\documentclass{aa}

\usepackage{graphicx}
\usepackage[varg]{txfonts}
\usepackage{natbib}
\bibpunct{(}{)}{;}{a}{}{,}
\usepackage{tikz}
\usepackage{caption,subcaption}
\usepackage{multirow}
\usepackage{hyperref}
\usepackage{gensymb}
\usepackage{lscape}
\usepackage{longtable}
\usepackage{diagbox}
\usepackage{supertabular,booktabs}
\usepackage{cleveref}
\usepackage{multicol}
\usepackage{amsmath}
\hypersetup{colorlinks=true, citecolor=blue, linkcolor=blue}
\definecolor{lightblue}{RGB}{51,131,255}

\begin{document} 

\renewcommand{\thefootnote}{\arabic{footnote}}

\title{Radial velocity homogeneous analysis of M dwarfs observed with HARPS. II. Detection limits and planetary occurrence statistics}

\titlerunning{RV Analysis of a M-Dwarfs H.A.R.P.S. sample. II.}

   \author{L. Mignon \inst{1,2}
   \and X. Delfosse \inst{1}
   \and N. Meunier \inst{1}
   \and G. Chaverot \inst{1,2,7}
   \and R. Burn \inst{8}
   \and X. Bonfils \inst{1}
   \and F. Bouchy \inst{2}
   \and N. Astudillo-Defru \inst{3}
   \and G. Lo Curto \inst{4}
   \and G. Gaisne \inst{1}
   \and S. Udry \inst{2}
   \and T. Forveille \inst{1}
   \and D. Segransan \inst{2}
   \and C. Lovis \inst{2}
   \and N. C. Santos \inst{5,6}
   \and M. Mayor \inst{2}
   }
   \authorrunning{L. Mignon et al.}

   \institute{Univ. Grenoble Alpes, CNRS, IPAG, F-38000 Grenoble, France   
\\ correspondence: \texttt{Lucile.Mignon@univ-grenoble-alpes.fr}
            \and Observatoire astronomique de l'Université de Genève, 51 chemin Pegasi, 1290 Versoix, Switzerland
            \and Departamento de Matemática y Física Aplicadas, Universidad Católica de la Santísima Concepción, Alonso de Rivera 2850, Concepción, Chile
            \and European Southern Observatory, Casilla 19001, Santiago 19, Chile
            \and Instituto de Astrofísica e Ciências do Espaço, Universidade do Porto, CAUP, Rua das Estrelas, 4150-762 Porto, Portugal
            \and Departamento de Física e Astronomia, Faculdade de Ciências, Universidade do Porto, Rua Campo Alegre, 4169-007 Porto, Portugal
            \and CVU - Life in the Universe Center, Geneva, Switzerland
            \and Max-Planck-Institut für Astronomie, Königstuhl 17, 69117, Heidelberg, Germany
   }

   \date{Received ; accepted }

 
  \abstract
  {}
   {We re-determine planetary occurrences around M dwarfs using 20 years of observations from HARPS on 197 targets. The first aim of this study is to propose more precise occurrence rates using the large volume of the sample but also variations to previous calculations, particularly by considering multiplicity, which is now an integral part of planetary occurrence calculations. The second aim is to exploit the extreme longevity of HARPS to determine occurrence rates in the unexplored domain of very long periods.}
   {This work relies entirely on the 197 radial velocity time series obtained and analysed in our previous study. By considering they are cleaned of any detectable signal, we convert them into detection limits. We use these 197 limits to produce a detectability map and combine it with confirmed planet detections to establish our occurrence rates. Finally, we also convert the detection limits from orbital period to insolation in order to construct an occurrence statistics for the temperate zone.}
   {We find a strong prevalence of low-mass planets around M dwarfs, with an occurrence rate of 120\% for planets with a mass between 0.75 and 3~M$_{\oplus}$. In addition, we compute an occurrence rate of $\rm 45.3^{+20}_{-16}\%$ for temperate zone planets around M dwarfs. We obtain an occurrence rate of a few percent for giant planets with wide separations. In our sample these giant planets with wide separations are only detected around the most massive M dwarfs.}
{}


   \maketitle


\section{Introduction}\label{intro}

M dwarfs represent 80\% of the Solar neighbourhood \cite[][]{henry1994,chabrier2000}, with more than 1500 stars within 26 pc, making them ideal targets for statistical analysis.
As the radial velocity (hereafter RV) signal of a planet strongly depends on the stellar-to-planet mass ratio, large surveys of M dwarfs have been monitored with RV planet searcher \citep[e.g. HARPS,][]{Mayor2003}.
Since the discovery of the first exoplanet orbiting an M dwarf \cite[][]{delfosse1998,marcy1998}, which was a massive planet, a total of 135 planets have been discovered around M dwarfs using RV technique\footnote{based on the NASA Exoplanet Archive \url{https://exoplanetarchive.ipac.caltech.edu/index.html}}. they span a wide range of masses, from 0.75 M$_{\oplus}$ \cite[GJ~54.1~b][]{astudillo2017B}, to 1570 M$_{\oplus}$ \cite[GJ~676A~b, 4.9 M$\rm _{J}$][]{forveille2011,anglada2012}.
The study of all these detections as a statistical set provides properties that individual detections and characterisations of planetary systems cannot offer, such as planetary and system occurrence rates.
These properties are derived from large samples, which can only be obtained with two detection methods today for short and intermediate periods: photometric transits and RVs.
The two Kepler transit missions yielded robust and detailed occurrence statistics of planets around FGK and M dwarfs based on orbital period and radii, especially for periods below 100 days \cite[e.g.][]{dressing2013,mulders2015,dressing2015,hsu2020}.
They predict an increase in occurrence frequency with decreasing star radius.
Concerning less massive planets (below 1~R$_{\oplus}$), the occurrence rates published vary between 50\% to 100\% for periods below 10 days and even more so between 10 and 100 days.
\cite{morton2014} and \cite{dressing2013} computed the number of planets per star of radii smaller than 4~R$_{\oplus}$ orbiting massive M dwarfs and find respectively 2~$\pm~0.45$ telluric planets and 0.9~$^{+0.04}_{-0.03}$ below respectively 150 and 50 days of period.

More recently, \cite{ment2023} established new occurrence rates of planets around very low mass M dwarfs, using the first 42 TESS sectors.
In their study, by comparing results with \cite{dressing2015}, they obtained a higher ratio of rocky to sub-Neptunes planets for low mass M dwarfs than for the massive ones.

Regarding the RV method, the extensive dataset of M dwarfs observed with HARPS has previously been the subject of a first occurrence analysis statistics established by \cite{bonfils2013a}, following the method previously proposed by \cite{cumming2008} and \cite{mayor2011} for a sample of FGK dwarfs.
Based at the time on a sample of 102 M dwarfs observed over more than six years with HARPS, they already announced the preponderance of short-period telluric planets around M dwarfs.
\cite{sabotta2021} established a new planetary occurrence statistics using a sample of 71 M dwarfs observed over five years with the CARMENES visible and infra-red spectrometers.
Despite the smaller star sample, this new analysis reaffirmed the dominance of short-period terrestrial planets reported.
More recently, in their study of 56 massive M dwarfs (M0 - M3) observed over six years with HARPS-N, \cite{pinamonti2022} reported a significantly lower occurrence rate compared to all previously cited studies.
They obtained only 0.10 planets per star for periods between 1 and 10 days.

Theoretical models of planet formation do so-far not offer a comprehensive interpretation of the increasing planet occurrence rate around low-mass stars. A lower occurrence of more massive planets around these objects can only partially explain the reported higher abundance of low-mass planets around low-mass stars \citep{burn2021}. However, detailed comparison of the same theoretical model to the CARMENES survey have not revealed a statistically significant under prediction of low-mass planets around M stars \citep{schlecker2022}.

The statistics that we present in this paper is built on a sample of 197 M dwarfs observed with HARPS over the last two decades.
Consequently, we delve into the long-period range that had remained unexplored by previous RV studies due to their limited temporal coverage.
This research is a continuation of the analysis of these 197 RV time series (\cite{mignon23b}, hereafter paper I), which allows us to propose here a substantial advancement in the methodology employed for constructing the occurrence statistics.
This study also aims to investigate the relationship between planetary population and the mass of the central star by comparing the populations obtained around the most and least massive M dwarfs in our sample.

The outline of this paper is the following. 
In Sect.~\ref{secSample}, we present the sample used to build the planetary occurrence statistics.
In Sect.~\ref{secDetec}, we describe the method used to compute the conservative and statistical detection limits. 
In Sect.~\ref{secStat} we present the implementation of the planetary and system occurrence statistics and discuss and compare the results in Sect.~\ref{result}.
We compute the occurrence in the temperate zone and present the results in Sect.\ref{TZ}.
Finally, we discuss the limits of this study in Sect.\ref{discu} and conclude in Sect.\ref{conclu}.


\section{Sample}\label{secSample}

This study focuses on 197\footnote{After detailled analysis, 3 targets are removed from the sample of 200 targets} M dwarfs selected from paper I, observed for at least 10 nights using the HARPS instrument between 2003 and 2021. We discuss the impact of this threshold in Sect.\ref{nmes-dependance}.
This sample is meticulously constructed by cross-referencing several catalogues of cool dwarfs\footnote{\cite{gaidos2014,winters2015,stauffer2010,henry2018,winters2021}} with the HARPS public archive of ESO\footnote{E2DS from \url{http://archive.eso.org/wdb/wdb/adp/phase3_spectral/form}} and considering only targets within a distance of less than 26 pc and below a declination of $ \rm +20^{\circ}\delta$.

\subsection{Magnitude and mass distributions}

We describe in paper I how we select VGJHK magnitudes from 6 different catalogues.
The absence of a consistent V-band magnitude catalogue for all M dwarfs in our sample necessitates employing only three out of the five empirical mass-luminosity relations from \cite{delfosse2000} (relations in the H, K, and J bands only) for stellar mass calculations.
As presented in paper I, the mass distribution in our sample ranges from 0.75 to 0.1~M$_{\odot}$ with an average (resp. median) value of 0.44 M$_{\odot}$ (resp. 0.46 M$_{\odot}$).

\subsection{Data}

All the programmes whose measurements are used in this study are listed in Table~\ref{progidtable}.
RV time series are extracted using a specific template-matching code designed for M dwarfs, previously established and used in multiple exoplanet detection studies \cite[e.g.][]{astudillo2015harps,astudillo2017A,astudillo2020}.
Following a sequence of successive selection steps, our sample is made up of 197 series comprising 10513 RVs, binned by night and corrected for secular acceleration.
For this second stage of analysis, we use exclusively the residual RV time series established in paper I to compute the detection limits and we do not correct them any more.

\subsection{Residuals time series of RV}

Regarding the long-term variability, 16 RV time series are corrected for a dominant linear model and 29 for a quadratic model. 
Once the long-term model is identified and corrected for, 36 RV time series are then corrected for one or multiple Keplerian signals either attributed to previously published planetary systems (listed in Table~\ref{tabledetection}), or arising as candidates (7 cases).
Finally, we corrected 23 RV time series for the identified signature of the rotation period,
which included not only those at the rotational period but also harmonics.
For exemple, in the case of GJ~581, we detected signals around 127 days in three chromospheric indices (Calcium, Sodium, $\rm H\alpha$), but we detected and subtracted a RV signal found at 67 days, which we attributed to the harmonic of the rotation period.
We employed a similar approach for GJ~447 (61.9 days and 123 days) and GJ~514 (15.15 days and 31.4 days).

\subsection{Final sample}
Ultimately, since the targets GJ~1132 and GJ~1214 were included in the blind search following the discovery of a planetary candidate by Mearth, we decided to exclude them from our sample to avoid biasing the statistics.
Additionally, identifying GJ~4254 as a spectroscopic binary, which prevents us from accurately isolating the signal from the primary companion, we excluded it.


\section{Detection limits}\label{secDetec}

We determine detection limits using the methodology outlined in \cite{bonfils2013a}, based on the Generalized Lomb Scargle (hereafter GLS) method proposed by \cite{zechmeister2009a}.
Given that all detectable signals within the RV time series are removed, we consider our residual RV time series as a realisation of white noise.

\subsection{Computation method}

Following the approach presented in \cite{bonfils2013a}, we first shuffle the RV values of a given star while keeping the original temporal sampling and then compute the GLS periodogram. This is performed 1000 times, and allows us to determine for each period the distribution of power that is compatible with no planet, i.e. due to noise. 
We then consider each period $P$ in the periodogram. 
We inject in the RV time series a planetary signal with a period of $P$, no eccentricity, and of a given semi-amplitude $K$: 12 realisations of equi-spaced phases are performed. We start with $K$ corresponding to the fitted sinusoid at $P$ in the original RV time series. The power $p$ in the periodogram at period $P$ is computed. $K$ is then progressively increased, until the fap level of the power of the peak down to 1\%. It corresponds to the power of the 990th values in the power distribution determined above, after ranking them in increasing order. The amplitude is then converted into a planetary mass (using the stellar mass estimated in our previous study). We therefore obtain, for each period $P$, 12 planetary mass estimations above which we can rule out the presence of the planet with a 1~\% confidence level. 
The amplitude of the signal is phase-dependent, and thus, we can choose either a highly conservative threshold by considering the worst phase and the largest amplitude (referred to as the conservative limit), or a statistical threshold by considering the median value (statistical limit). In this statistical study, we adopt the same assumption as \cite{bonfils2013a}, and exclusively use the statistical detection limits. This assumes that it is highly improbable for all systems to be observed under the most unfavorable phase conditions for detection

\begin{figure}
   \centering         
    \includegraphics[width=0.49\textwidth]{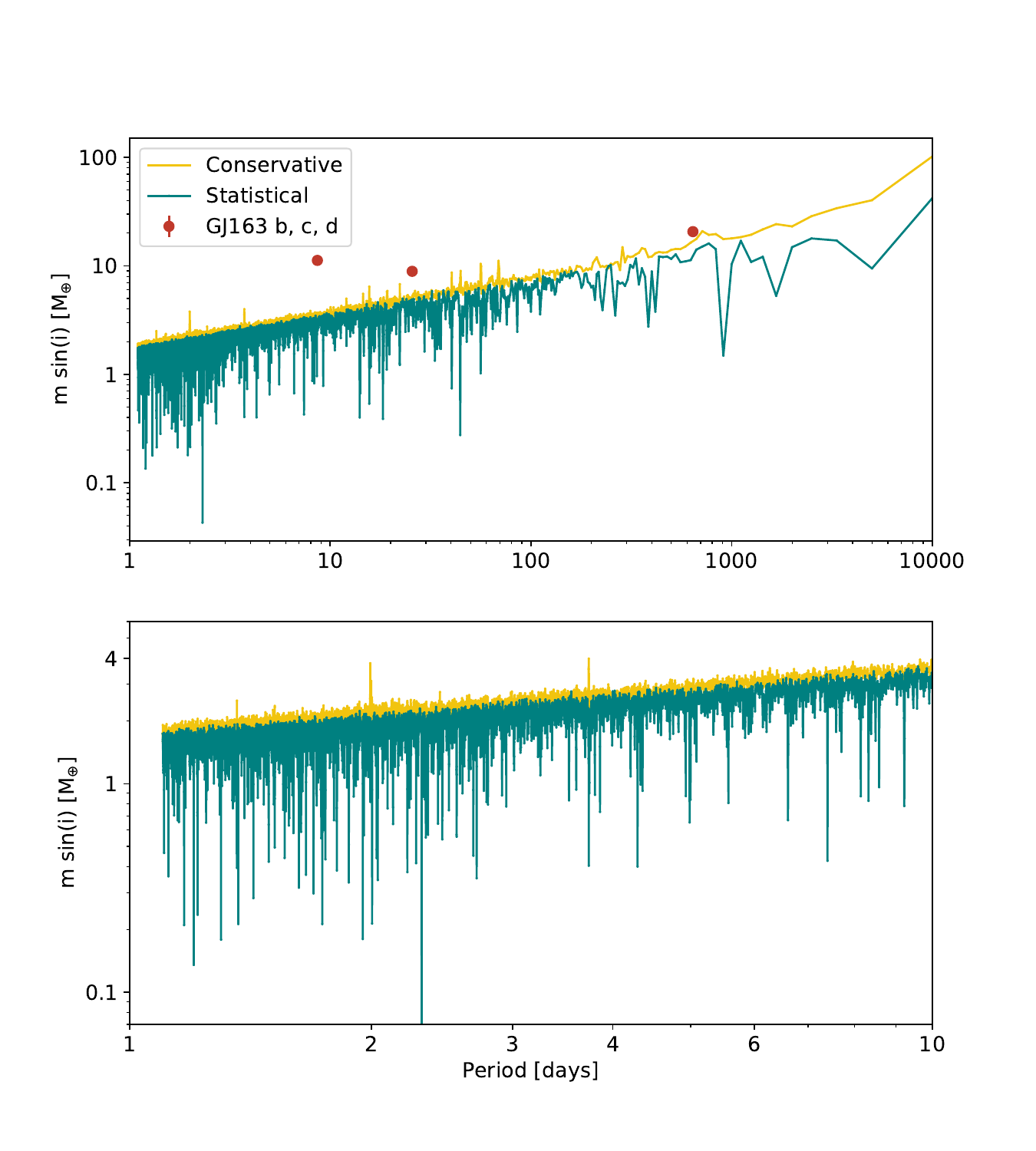}
    \caption{Conservative (yellow) and statistical (dark turquoise) detection limits of GJ~163 as function of projected planetary mass and periods, with the respective positions of the detections of GJ~163~b, c and d. Upper panel shows the limits in a range of periods between 1.6 and 10000 days, while the lower panel shows a zoom of these limits for periods between 1.6 and 10 days.}
    \label{limdet-gj163}
\end{figure}

This is done for periods between 1.6 and 10000 days and the 197 stars in our sample.
As an example, Fig.~\ref{limdet-gj163} shows the statistical (dark turquoise) and conservative (yellow) detection limits for GJ~163 as a function of the period.

It is worth noting that this methodology incorporates the injection of sinusoidal functions, not Keplerian ones, into the RV series.
According to \cite{cumming2010}, this modeling remains robust up to an eccentricity of 0.5, suggesting that any signal with a higher eccentricity would be underestimated.
However, considering that 81.4\% of the detected planets possess an eccentricity lower than 0.4\footnote{according to the NASA exoplanet database}, it allows us to consider this modelling as sufficiently conservative in this statistical study.
Furthermore, this method is correct under the assumption that the noise on the data is white.
We discuss the consequences of this in detail in section 6.2.

\subsection{Detection map}

\begin{figure}
   \centering         
    \includegraphics[width=0.49\textwidth]{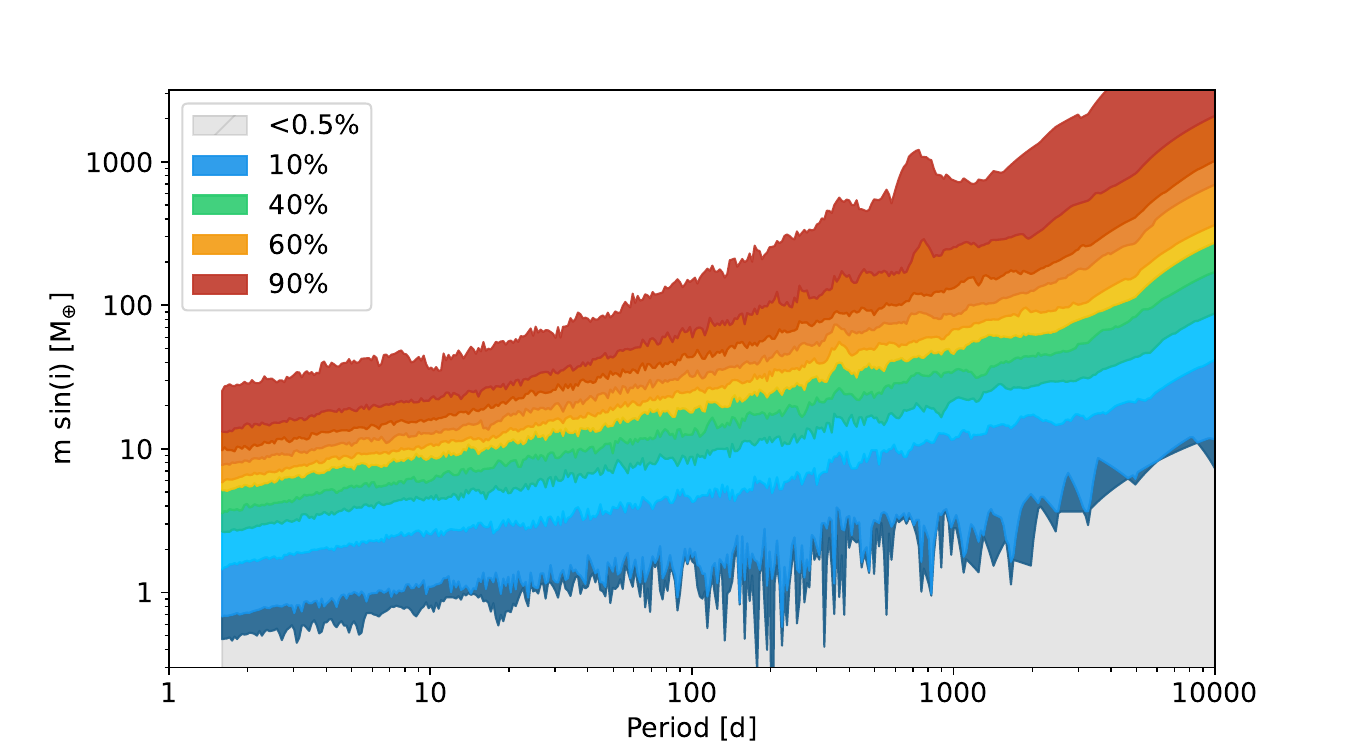}
    \caption{Iso-frequency map obtained by stacking the 197 statistical detection limits in projected masses as a function of the period. Iso-contours are highlighted for several percentage levels of the sample and describe the distribution of maximum mass missed. The grey part represents the unknown domain, below the lowest statistical detection limit (corresponding to 0.5\%), and the white part is the domain beyond 100\%.}
    \label{limdet}
\end{figure}

The linear frequency sampling of the GLS periodogram makes the detection limits extremely erratic and noisy at short periods (as illustred on the lower plot of Fig.~\ref{limdet-gj163}).
As detailed by \cite{zechmeister2009a}, the frequency step corresponds to the inverse of the longest tested period, which in our case is the same for all targets: 10,000 days. This means that very high frequencies (corresponding to very short periods) are tested with inadequate sampling (e.g., 11 nights over a single season), leading to oversampling these short periods, especially compared to the uncertainties associated with the fitted planetary periods.
To mitigate erratic rates, we smooth these limits over the period range of 1.6 to 10 days (400 points sampled on a log-uniform scale).
We then generate the detection map for our dataset by stacking the statistical detection limits\footnote{The full map is composed of 1000 mass values ranging from log-0.5 to
log3.5 $\rm M_{\oplus}$ on a log uniform distribution, and 1400 period values ranging from 1.6
to 10000 days on a log uniform distribution after the smooth operation.}.
Figure.~\ref{limdet} illustrates the distribution of detection limits of projected mass as a function of the probed orbital period.
In this representation, the color-coded scheme enhances the visualization of iso-frequency curves (the fraction of
stars on our sample around which a given planet could be detected).
The grey area at the bottom of the map corresponds to the out-of-reach domain (below the lowest detection limit of 0.5\%).
This region encompasses planets with projected masses ranging from 0.6 to 1~M$_{\oplus}$ for periods under 10 days, and extending up to 20~M$_{\oplus}$ for periods as long as 1000 days.
Conversely, the white area in the upper-left corner of the map corresponds to the domain where planets cannot be missed (beyond the highest detection limit) with a 99\% confidence level.
This regions encompasses planets with projected masses ranging from 20 to 50~M$_{\oplus}$ for periods under 10 days, up to 1000~M$_{\oplus}$ for periods under 1000 days.
The iso-frequency contours on this map are interpreted as follows: if planets in the green zone exist around M dwarfs in this sample, then we should expect to detect them in 30\% to 40\% of cases (around 60 to 80 stars among our 197) in this specific region.

Finally, this gives the local detectability metric for the mass-period map, which will be used to weight each planet detection in the occurrence rate calculations and to characterize the accessibility of any size of domain in the next section.


\section{Construction of the occurrence statistic}\label{secStat}

Planetary occurrence rates provide the expected count of planets around stars, considering both the detected planets and the detection limits.
Consequently, a rate of 100\% implies for example an expectation of 100 planets orbiting 100 stars, without specific knowledge about their distribution within the systems.
We partition the map of projected planetary mass and orbital period (hereafter referred to as the mass-period map) into subdomains where we compute these occurrence rates.
In the subsequent sections of this study, a domain $jk$ therefore refers to a region on this map, delimited by projected mass values between $m_k$ and $m_{k+1}$ and period values between $p_j$ and $p_{j+1}$.

\subsection{Detections}\label{detection}

In \cite{bonfils2013a}, for a given domain, occurrence rate was the ratio of the number of detections $ndet$ to the number of accessible targets $N_{eff}$.
To take into account the variation in the detectability of a planet within a domain, in this work we adopt another method to compute the occurrence rate of a given domain: the ratio of the number of predicted planets in the domain to the total number of targets in the sample.
In the $jk$ domain of the mass-period map, we compute the local occurrence frequency $f_{jk}$ as:

\begin{equation}
    \rm f_{jk} = \frac{\widetilde{n}_{jk}}{N} = \frac{1}{N} \times \sum_{i=1}^{n_{jk}} {\alpha_{(m_{i};p_i)}}^{-1} 
\label{occu}
\end{equation}

where $\alpha_{(m_i;p_i)}$ is the local detectability rate of the planet $i$, $n_{jk}$ is the total count of detections in the $jk$ domain and $N$ is the total number of stars of the sample.
$\widetilde{n}_{jk}$ is therefore the weighted number of detections in the domain.
As an illustration, a unique detection with a local detectability rate of about 50\%, on a sample of 100 stars, gives an occurrence rate of 2\%.

The two ratios give identical results under two circumstances: when detections are uniformly distributed within the domain, or when the domain's detectability rate remains constant (which only occurs in the left top corner and in the right bottom corner).
Otherwise, rates obtained with the ratio used in \cite{bonfils2013a} can be over-estimated if detections are in the top of the domain whereas the number of accessible targets is computed for a average mass and period in the range in the domain.
Conversely, these rates can also be under-estimated if detections are located in the least massive planets parts of the domain.
As a result, occurrence rates are very sensitive to the position of the detected planet in the mass-period map.
We also note that they can be over 100\%, which implies local multiplicity.

\subsection{Coefficient of multiplicity}

In a general case, we compute the statistical bounds of the occurrence rates using a binomial law: we consider a detection as a success of the event. 
This law admits a maximum frequency of 100\% for N planets detected around N stars.
This is why multiplicity plays an important role in this work.
The local coefficient multiplicity $\rm mult_{jk}$ is the average multiplicity of all the stars accessible (with and without detection) in the $jk$ domain.
This definition of local multiplicity in a domain combines observed multiplicity and probability of additionnal planets.
We compute the multiplicity $\rm mult_{i\:jk}$ of each star in the $jk$ domain as:

\begin{equation}
	\rm mult_{i\:jk} = mult_{obs\:i\:jk} + \overline{\alpha_{\textit{jk}}} = mult_{obs\:i\:jk} + 1 ~ - < \alpha >_{\textit{jk}} 
\label{multjk}
\end{equation}

where $mult_{obs\:i\:jk}$ is the observed multiplicity of the system i in the $jk$ domain, $< \alpha >_{jk}$ is the average detectability rate in the domain, which is a probability of detection, and therefore, $\overline{\alpha_{jk}}$ is the probability of a missed planet in a system in the considered $jk$ domain.
This coefficient gives a more complex multiplicity than the observed one (average number of planets in each systems in the domain) because it takes into account the limit of detectability of the domain.
For a given system of 1 planet, if the local detectability rate is around 100\% (as in the region corresponding to massive planets at short period for example), the weighted multiplicity is only 1 because we know we can not miss a detection with a confidence level beyond 99\%. 
Conversely, if the local detectability rate of the domain is 60\% for example, there is a non negligible probability to have missed a second planet, therefore the weighted multiplicity is higher than 1 for this system.

\subsection{Statistical bounds}

We compute statistical bounds by using a binomial law of N pickings with replacements because events are not linked to each other and can only have two possible outcomes, i.e. detection or no detection.
As the occurrence rate can reach 100\% and beyond, we use the multiplicity coefficient to compute the statistical bounds.
Returning to a given $jk$ domain, with a $mult$ multiplicity, a $f$ occurrence frequency and $n$ detections, we compute the 1$\sigma$ statistical bounds (lower $f_{low}$ and upper $f_{up}$) as:

\begin{subequations}
	\begin{align}
\int_{0}^{f_{up}} \left(_{N_{eff}\times mult}^{n} \right)~\times~f^{n}~(1-f)^{(N_{eff} \times mult-n)}~ df = 0.84 \\
\int_{f_{low}}^{1} \left(_{N_{eff}\times mult}^{n} \right)~\times~f^{n}~(1-f)^{(N_{eff} \times mult-n)}~ df = 0.84 
\end{align}
\label{bornefinale}
\end{subequations}

where $N_{eff}$ is the number of accessible stars in the domain used in \cite{bonfils2013a}.
We obtain $N_{eff}$ from multiplying the total number of stars within the sample by the average probability of detection in the domain (obtained by a log-uniform Monte-Carlo sampling of 15000 trials in the domain).
This computation assumes a log-uniform probability distribution in both mass and period, as in \cite{bonfils2013a}.
Finally, the greyed-out part of Fig.~\ref{limdet} is not included in the subsequent calculations.

Through this computation, we establish the $\rm 1\sigma$ confidence interval for the binomial probability distribution associated with achieving $n$ successes in $N_{eff}.mult$ trials with a $f$ probability of success.
In domains where multiplicity computed is below 1, we consider only one trial per accessible target, i.e. $\rm mult$ parameter set to 1 in Eq.~\ref{bornefinale}.
Differing from previous studies, we consider multiplicity, enabling the possibility of drawing a star anew, irrespective of the outcome of the previous draw.
This adaptation stems from the fact that the binomial distribution implies independent random draws.
However, it is important to acknowledge a methodological limitations here: within a given domain, the detection of a subsequent planet is not entirely independent from the initial planet detection.
This limitation is discussed in Sect.\ref{discu}.
For domains without detection, we exclusively compute an upper limit through the construction of a binomial distribution assuming zero successes within a number of trials equivalent to the number of accessible stars in the domain $N_{eff}$.

\subsection{Planetary system occurrence statistic}\label{subsecStatsyst}

Beyond the number of expected planets within a sample of 100 stars, this study also examines the occurrence of planetary systems, namely the number of expected planetary systems per 100 examined stars.
It is crucial to ascertain whether an occurrence rate of 100 planets among 100 stars corresponds to 100 single systems or 50 multiple systems for example.
For this computation, we count the number of systems rather than the number of individual planets.
Within each domain, a multiple system is counted only once, and the local weight attributed to the system is the individual weight of the first planet (easiest to detect) composing the system.
The occurrence rate of planetary systems within a domain is therefore ratio similar to the planetary occurrence.
Consequently, the statistical bounds are determined analogously to the previous case, but with a number of detection corresponding to the number of the systems and a multiplicity value of 1.
This computation is a first estimation of the system occurrence rates, according to the method of computation we chose.


\section{Results}\label{result}

In this section, we present the planetary and system occurrence rates obtained through the method described above.
Additionally, we provide the rates derived from sub-samples, which represent distinct populations aimed at better understanding the relationship between planetary populations and the physical parameters of the central star.

\subsection{Planetary occurrence rates}

\begin{figure*}
   \centering         
    \includegraphics[width=0.8\textwidth]{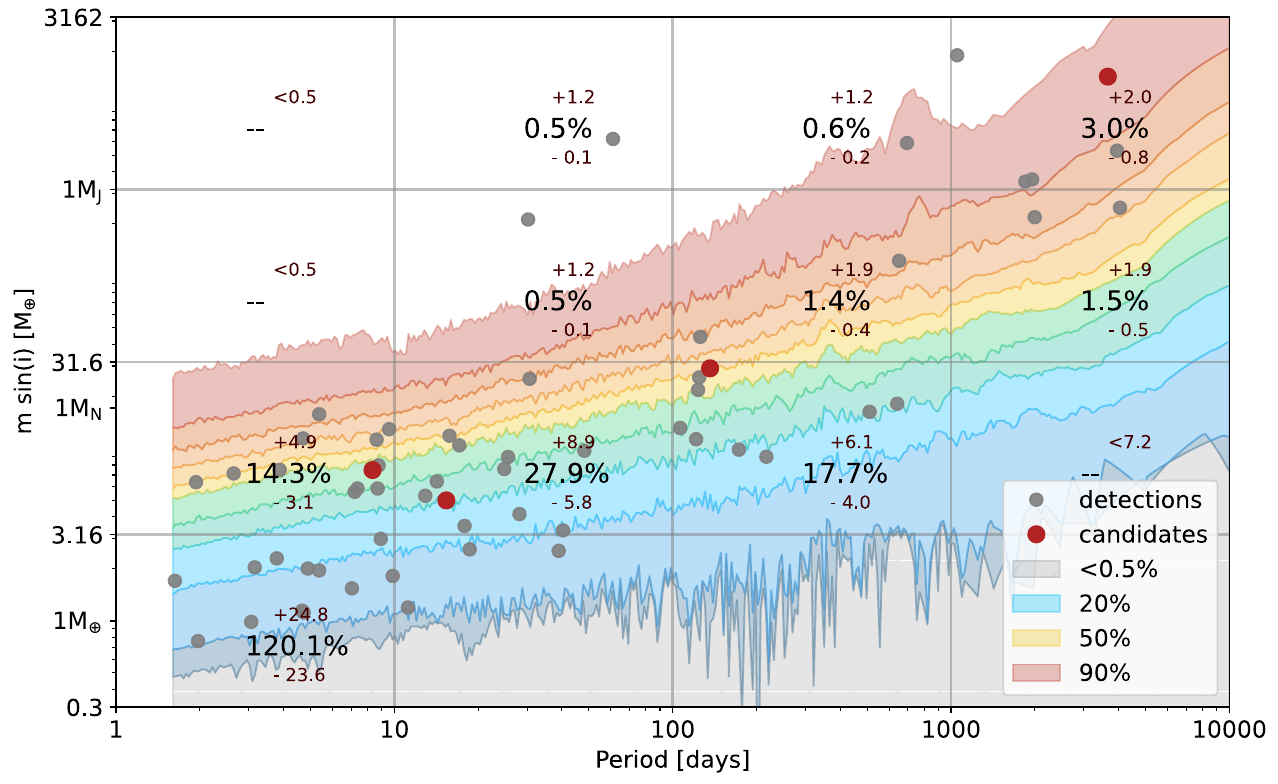}
    \caption{Planetary occurrence statistic. Colours in the background refer to the detection limit map, from the non-reachable region (grey) to the unmissable one (white). The grey points represent the considered detections while the red points represent the candidates. The grey horizontal and vertical lines delineate the probed domains. The computed occurrence rates (along with their corresponding statistical bounds) are indicated at the center of each domain.}
    \label{statocc}
\end{figure*}

To initiate the presentation of results, Fig.~\ref{statocc} presents the rates from the entire sample of 197 M dwarfs as a function of projected planetary mass and orbital period.
Within this visual representation, we overlay the detectability map and the demarcations of the previously discussed domains, in addition with the occurrence rates and their associated statistical bounds.
For this first representation, we categorise the planetary masses into four ranges: terrestrial planets of very low mass (\textit{m} sin \textit{i} $<$ 3.16 $\rm M_{\oplus}$), intermediate-mass planets ranging between super-Earths and Neptunes (3.2 $<$ \textit{m} sin \textit{i} $<$ 31.6 $\rm M_{\oplus}$), intermediate-mass of volatile planets (31.6 $<$ \textit{m} sin \textit{i} $<$ 1 $\rm M_J$) and Jovian planets (\textit{m} sin \textit{i} $>$ 1 $\rm M_{J}$).
A projected mass of 3162 Earth masses corresponds statistically to an actual mass of approximately 4700 Earth masses, which exceeds the 13 Jupiter mass threshold considered as the boundary between massive planets and brown dwarfs by the IAU (deuterium fusion critical mass).
This is why we choose not to explore the higher mass range and propose this segmentation.

In the lower mass range (\textit{m} sin \textit{i} $<$ 3.16 $\rm M_{\oplus}$), we obtain an occurrence frequency of $\rm 120\%^{+24.8}_{-23.7}$ for periods shorter than 10 days.
Considering that the longer-period domains are poorly covered by the detection map, we do not compute the frequencies for these planets beyond 10 days.
In the higher mass range, including Super-Earth and Neptunes, we obtain a rate of $\rm 14.3\%^{+4.8}_{-3.1}$ for periods shorter than 10 days, $\rm 27.9\%^{+8.9}_{-5.8}$ between 10 and 100 days, $\rm 17.7\%^{+6.1}_{-4.1}$ between 100 and 1000 days, and finally, an upper limit of 7.2\% beyond 1000 days as we have no detections in this less explored domain.

Table~\ref{plvssyst} presents both planetary occurrence and system frequencies within these two specified mass ranges at short orbital periods.
Within the lower mass range, at periods shorter than 10 days the expected $\rm 120\%^{+25}_{-24}$ planetary occurrence frequency is concentrated around $\rm 80\%^{+16}_{-18}$ of the targets accessible in the domain.
These frequencies could indicate that 20\% of the systems investigated with HARPS do not host planets of extremely low mass with short orbital periods.
Conversely, the remaining systems would occasionally exhibit the presence of multiple planets, which would necessarily implies the existence of distinct and varied planetary systems within the surveyed sample.
This hypothesis is reinforced by the multiplicity of 1.6 computed in this domain.
Within the higher mass ranges, the expected quantity of planets is fully compatible with the expected number of systems, thus suggesting no multiplicity in the other domains.

\begin{table}
\caption{Planetary and systems occurrence rates}
\label{plvssyst}
\begin{center}
\renewcommand{\footnoterule}{}  
    \begin{tabular}{ l | c  c | c  c }
      & \multicolumn{2}{c }{1.6 - 10~d} & \multicolumn{2}{c}{10 - 100~d} \\
     & $\rm f_{p}$ &  $\rm f_{s}$ &  $\rm f_{p}$ &  $\rm f_{s}$ \\ \hline
     &&&& \\ 
    $\rm 3.2 - 31.6~M_{\oplus}$ & $\rm 14^{+5}_{-3}$ & $\rm 14^{+5}_{-3}$ & $\rm 28^{+9}_{-6}$ & $\rm 27^{+9}_{-6}$ \\
    &&&& \\
    $\rm < 3.2~M_{\oplus}$ & $\rm 120^{+25}_{-24}$ & $\rm 80^{+16}_{-18}$ & - & -  \\
    
\end{tabular}
\end{center}
\tablefoot{Comparison between planetary occurrence rates ($\rm f_{p}$) and systems occurrence rates ($\rm f_{s}$), in percent.}
\end{table}

In the intermediate volatile planet mass range (31.6 $<$ \textit{m} sin \textit{i} $<$ 1 $\rm M_J$), we obtain an upper limit of 0.5\% for periods shorter than 10 days because we have no detection in this domain, even though it is well constrained by detection limits.
This absence of planets in this mass range at short periods is not a bias of the study or observational, but rather an absence because such planets could not have been missed at such a period.
Between 10 and 100 days, we obtain a rate of $\rm 0.5\%^{+1.2}_{-0.1}$, then $\rm 1.4\%^{+1.9}_{-0.4}$ between 100 and 1000 days, and finally $\rm 1.5\%^{+1.8}_{-1.5}$ between 1000 and 10000 days.
In the mass range of Jovian planets (\textit{m} sin \textit{i} $>$ 1 $\rm M_{J}$), we again obtain an upper limit for periods shorter than 10 days because we have no detection in this domain, despite being well constrained.
We obtain a rate of $\rm 0.5\%^{+1.2}_{-0.1}$ between 10 and 100 days, then $\rm 0.6\%^{+1.2}_{-0.2}$ between 100 and 1000 days , and finally $\rm 3.0\%^{+2.0}_{-0.8}$ between 1000 and 10000 days.

Table~\ref{tableoccu_value} lists all planetary occurrence rates $f_p$ computed and presented in Fig.~\ref{statocc}.
It also provides the number of detections $n_d$ and the number of accessible targets in the domain $N_{eff}$.
In addition, the parameter $N_{eff'}$ provides a clearer representation of the difference between the calculation based on the average weighting of the domain and the approach used in this study, which applies local weighting to each detection.
The two calculation are equivalent if detections are homogeneously distributed in the domain, when $N_{eff}$ is equal to $N_{eff'}$.
When $N_{eff}$ > $N_{eff'}$, the detections are concentrated in the upper part of the surveyed domain, and conversely, when $N_{eff}$ < $N_{eff'}$, the detections are located in the lower part of the domain.

\begin{table*}
\caption{\label{tableoccu_value}Occurrence rates}

\begin{center}
\renewcommand{\footnoterule}{}  
\begin{tabular}{   l | c || c | c|  c|  c}

Planetary mass range  & Occurrence & \multicolumn{4}{c }{Period range [d]} \\
$[\rm M_{\oplus}]$ & Parameters & 1.6 - 10 & 10 - 100 & 100 - 1000 & 1000 - 10000 \\
\hline
\hline

\multirow{4}{6em}{316 - 3162} & $f_p$ & $< 0.5 \%$  & $\rm 0.5^{+1.2}_{-0.1} \% $ & $\rm 0.6^{+1.2}_{-0.2} \% $ & $\rm 3.0^{+2.0}_{-0.8} \% $ \\
& $n_d$&  0 & 1 & 1 & 5 \\
&  $N_{eff'}$ & - & 190 & 175  & 164 \\
&  $N_{eff}$& 197 & 191 & 185 & 162 \\ \hline

\multirow{4}{6em}{31.6 - 316} & $f_p$  & $< 0.5 \%$  & $\rm 0.5^{+1.2}_{-0.1} \% $ & $\rm 1.4^{+1.9}_{-0.4} \% $ & $\rm 1.5^{+1.9}_{-0.5} \% $ \\
& $n_d$&  0 & 1 & 2 & 2 \\
&  $N_{eff'}$ & - & 189 & 137  & 132 \\
&  $N_{eff}$& 188 & 178 & 141 & 86 \\ \hline

\multirow{4}{6em}{3.2 - 31.6} & $f_p$  & $\rm 14.5^{+4.9}_{-3.1} \% $ & $\rm 27.9^{+8.9}_{-5.8} \% $ & $\rm 17.7^{+6.1}_{-4.0} \% $ & - \\
& $n_d$&  12 & 12 & 8 & - \\
&  $N_{eff'}$ & 83 & 42 & 52  & - \\
&  $N_{eff}$& 115 & 73 & 35 & 13 \\ \hline

\multirow{4}{6em}{0.3 - 3.2} & $f_p$  & $\rm 120.1^{+24.8}_{-23.6} \% $ & - & - & -  \\
& $n_d$&  10 & - & - & - \\
&  $N_{eff'}$ & 8 & - & -  & - \\
&  $N_{eff}$& 14 & - & - & - \\

\end{tabular}
\end{center}
\tablefoot{Planetary occurrences rates ($f_p$) in percent as function of period and planetary mass domain probed. $n_d$ is the number of detection in the domain where $N_{eff}$ is the number of accessible targets. }
\end{table*}

\subsection{Dependence on the central mass}

In this section, we examine the correlation between the planetary population and the mass of the central star.
For this purpose, we split the sample according to the mass of the central star: those below and above 0.35 solar masses, limit between partially and fully convective \citep{chabrier1997}.
Therefore, we are analyzing and comparing two distinct samples: the first comprises 145 very massive M dwarfs, with a median stellar mass of $\rm 0.48~M_{\odot}$, while the second consists of 52 less massive M dwarfs, with a median stellar mass of $\rm 0.26~M_{\odot}$.

\begin{figure}
   \centering         
    \includegraphics[width=0.49\textwidth]{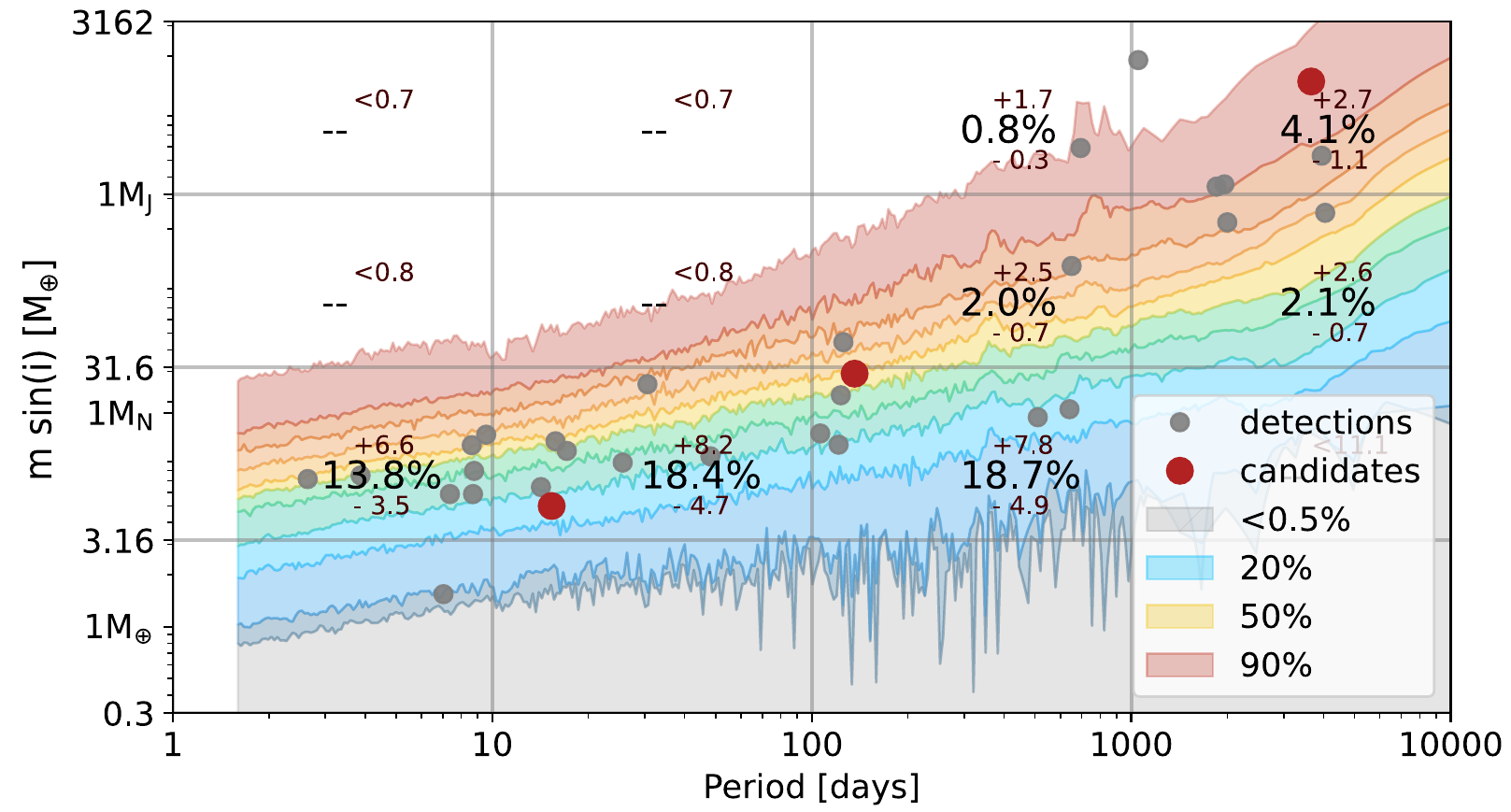}
    \caption{Planetary occurrence statistics obtained on the sub-sample of M-dwarfs with $\rm M_* > 0.35~M_{\odot}$}
    \label{statocc_sup35}
\end{figure}

\begin{figure}
   \centering         
    \includegraphics[width=0.49\textwidth]{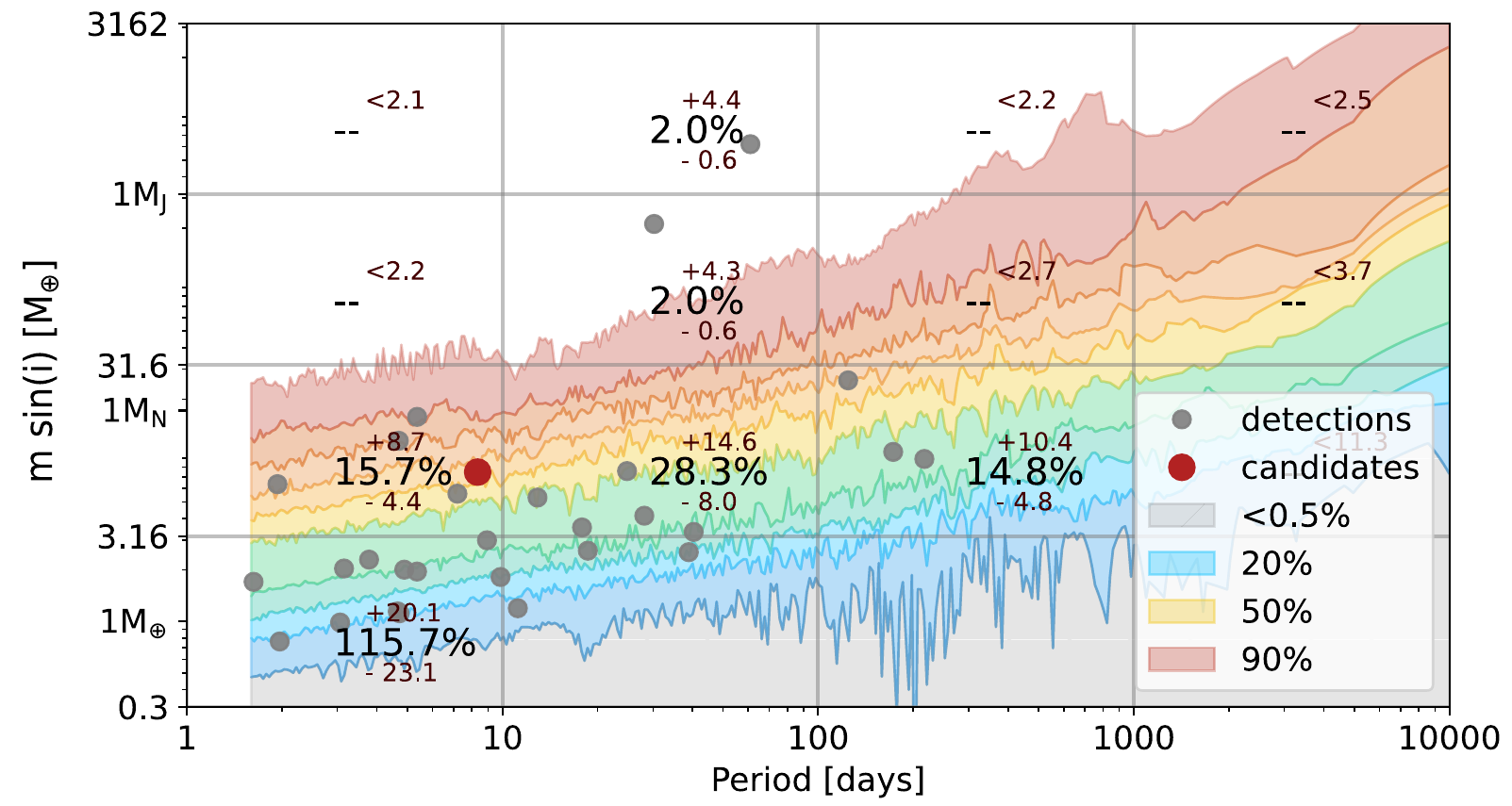}
    \caption{Planetary occurrence statistics obtained on the sub-sample of M-dwarfs with $\rm M_* < 0.35~M_{\odot}$}
    \label{statocc_inf35}
\end{figure}

In order to compare the two planetary populations, we calculate the rates within the same four mass ranges: terrestrial planets of very low mass (\textit{m} sin \textit{i} $<$ 3.16 $\rm M_{\oplus}$), intermediate-mass planets ranging between super-Earths and Neptunes (3.2 $<$ \textit{m} sin \textit{i} $<$ 31.6 $\rm M_{\oplus}$), intermediate-mass of volatile planets (31.6 $<$ \textit{m} sin \textit{i} $<$ 1 $\rm M_J$), and massive Jovian planets (\textit{m} sin \textit{i} $>$ 1 $\rm M_{J}$).
Figure~\ref{statocc_sup35} displays the occurrence rates obtained for the sub-sample of more massive M dwarfs ($\rm M_* > 0.35~M_{\odot}$) and Fig~\ref{statocc_inf35} shows those obtained for the sub-sample consisting of the less-massive ones ($\rm M_* < 0.35~M_{\odot}$).

\begin{table*}
\caption{\label{tableoccu_masse} Occurrence rates dependence on the central mass.}

\begin{center}
\renewcommand{\footnoterule}{}  
\begin{tabular}{   l | c | c || c | c|  c|  c}

Planetary mass range  & Sample & Occurrence f & \multicolumn{4}{c }{Period range [d]} \\
$[\rm M_{\oplus}]$ & [$\rm M_{\odot}$]  & Detections $n_d$& 1.6 - 10 & 10 - 100 & 100 - 1000 & 1000 - 10000 \\
\hline
\hline

\multirow{4}{6em}{100 - 3162} & \multirow{2}{*}{$\rm M_* < 0.35$} & f & < 2 \% & $\rm 4^{+5}_{-1} \% $ & < 2 \%  & < 3 \%  \\
& & $n_d$&  0 & 2 & 0 & 0 \\ \cline{2-7}
& \multirow{2}{*}{$\rm M_* > 0.35$} & f & < 0.7 \% & < 0.7 \% & $\rm 2^{+2}_{-1} \% $ & $\rm 6^{+3}_{-2} \% $ \\
& & $n_d$& 0 & 0 & 2 & 7 \\ \hline

\multirow{4}{6em}{3.2 - 100} & \multirow{2}{*}{ $\rm M_* < 0.35$} & f  & $\rm 16^{+9}_{-4} \% $ & $\rm 28^{+15}_{-8} \% $ & $\rm 15^{+11}_{-5} \% $ & < 7 \% \\
& & $n_d$&  5 & 5 & 3 & 0 \\ \cline{2-7}
& \multirow{2}{*}{$\rm M_* > 0.35$} & f & $\rm 14^{+7}_{-4} \% $ & $\rm 18^{+9}_{-5} \% $& $\rm 20^{+9}_{-5}\% $  & < 5 \% \\
& & $n_d$& 7 & 7 & 7 & 0 \\ \hline

\multirow{4}{6em}{< 3.2} & \multirow{2}{*}{ $\rm M_* < 0.35$} & f & $\rm 115^{+20}_{-23} \% $ & - & - & - \\
& & $n_d$&  10 & - & - & - \\ \cline{2-7}
& \multirow{2}{*}{$\rm M_* > 0.35$} & f & $\mathit{100^{+153}_{-36} \% }$ & - & -  & - \\
& & $n_d$& \textit{1} & - & - & - \\

\end{tabular}
\end{center}
\tablefoot{Comparison between planetary occurrence rates in percent obtained on the sample of 145 massive M dwarfs ($M_*>0.35~M_{\odot}$) and the sample of the 52 least massive ones ($M_*<0.35~M_{\odot}$)}
\end{table*}

Regarding the smallest planetary mass range, the majority of detected planets orbit least massive M dwarfs, resulting in an occurrence rate of $\rm 116\%^{+20}_{-23}$ for periods shorter than 10 days for those stars.
This rate is based on a total of 11 detections distributed among 8 different systems in a domain where only 11 stars are accessible.
Concerning the more massive M dwarfs, we count a unique detection (GJ~393~b) only, in a domain that is poorly covered by the detectability map and consequently we do not calculate the occurence rate in this mass range domain for the sample of more massive M dwarfs.
However, this unique detection occurs in a domain where about 10\% of the systems are accessible, suggesting a lower occurrence rate compared less massive stars.
This reinforces the hypothesis that the occurrence frequency of these small terrestrial planets increases as stellar mass decreases (e.g., \cite{ment2023}).

In the intermediate planetary mass range (Super-earth to Neptunes), the occurrence rates are fairly homogeneous and compatible up to 1000 period days for the two sub-samples of M dwarfs.
Beyond 1000 days, no planets of this mass range have been detected in the sub-samples, resulting in two upper limits of 4.8\% and 7.4\% for the more massive and less massive M dwarfs, respectively, but this domain is poorly covered for both stellar population.

Finally, for massive planets (from Neptune to the brown dwarfs limit), there is a hint of a dichotomy between the two planetary populations.
Specifically, the only two giant planets detected at short periods (<100 days) orbit low-mass M dwarfs (both orbiting the same system GJ~876), while all giant planets at long periods orbit the more massive M dwarfs (upper right corner of Fig.~\ref{statocc_sup35}).
This difference in the population of giant planets at large separations is discussed in Sect.~\ref{discu}.

\subsection{Comparison and evolution}

We now compare the results obtained with this new calculation method and on this new HARPS sample with key studies.
In order to facilitate a meaningful comparison of these rates with our study, we adopt here the same mass segmentation and represent the occurrence rates obtained by the studies in Fig.~\ref{comp-bonfils} and we reported them in Tab.~\ref{tableoccu_comp} for each probed domain.
First, we compare with the occurrence rates obtained by \cite{bonfils2013a}, ten years ago, on a sample of M dwarfs observed with HARPS.
Then, we propose to compare our results with the occurrence rates obtained more recently by \cite{sabotta2021} on a sample of M dwarfs observed with CARMENES, and finally, we compare the results obtained on the sub-sample of the most massive M dwarfs to the recent study of \cite{pinamonti2022}.

\subsubsection{From 2013 to 2023}

The detection rates reported by \cite{bonfils2013a} are based on more than 2000 RV measurements from 102 targets and resulted in 20 detections, including 11 new ones.
None of these detections are massive planets (\textit{m} sin \textit{i} $>$ 100 $\rm M_{\oplus}$), leading to an upper limit of 1\% of hot Jupiters (period < 10~d).
As this study is based on a maximum temporal coverage of 7 years, the occurrence rates of massive objects with very long periods were not extensively explored (constrained by detection limits in this regime).

Concerning planets in the smallest range of mass that could be probed (1 $<$ \textit{m} sin \textit{i} $<$ 10 $\rm M_{\oplus}$), \citet{bonfils2013a} predicts occurrence rates of 36\%~$^{+24}_{-10}$ for periods between 1 and 10 days, and 52\%~$^{+50}_{-16}$ for periods between 10 and 100 days.
In this smallest mass range, we computed a frequency of 89\%~$^{+18}_{-15}$ incompatible at a 1-$\sigma$ level for very short periods (p < 10 days).
In this domain, there were 5 detections out of 14 accessible stars ($\rm N_{eff}$) in the previous study, while in our study, we have 18 detections out of 48 accessible stars.
Within this domain, the detectability rate ranges from 0.5\% to 70\% in our study (see Fig.~\ref{limdet}), and between 1\% and 50\% in \cite{bonfils2013a}.
Consequently, the incompatibility is due to the new computation method we use in this study, and it highlights the consequences of the assumptions made in both methods, as described in Sect.~\ref{detection}.

This situation is also observed for the mass range between 10 and $\rm 100~M_{\oplus}$ and periods between 100 and 1000 days.
The upper bound obtained in \cite{bonfils2013a} is incompatible with the rate obtained in this study, based on 8 detections inhomogeneously distributed in a domain where detectability rates vary significantly.
In the other domains, the rates obtained by both studies are in agreement (within a 1-$\sigma$ confidence interval), but enlarging the sample size improves uncertainties.
Importantly, this new study is also built upon a larger number of detections, ultimately allowing for precise rates to be obtained in five domains where only an upper-bound estimate was previously possible without detections.
Finally, this mass segmentation allows to probe between $10^{3}$ and $\rm 10^{4}~M_{\oplus}$, however, it should be noted that we have not included in our analysis objects larger than $\rm 13~M_J$ (i.e. approximately $4100~M_{\oplus}$).
Although this mass range extends beyond this limit, any potential detections therein are not included, and thus, the rates obtained are underestimated.

\subsubsection{CARMENES occurrence rates obtained in RV}

\cite{sabotta2021} (hereafter S21) built their study on the 71 M dwarfs observed at least 50 nights from the 329 M dwarfs followed in the CARMENES survey.
The occurrence rates are based on 27 detections corresponding to 21 planetary systems.
As a main result, they computed more than one small planet (1 $<$ \textit{m} sin \textit{i} $<$ 10 $\rm M_{\oplus}$) for periods shorter than 100 days ($\rm 132\%^{+33}_{-32}$).
As can be seen in the Tab.~\ref{tableoccu_comp}, the rates obtained from our sample are all consistent (within $1-\sigma$) with those from the CARMENES study, as well as with the upper bounds in the detection-lacking domains.

Regarding the stellar mass dependence, we did indeed conduct our calculations in the same domains to facilitate comparison, although we did not include this in the paper.
S21 chose to set the stellar mass cut-off at $0.34~M_{\odot}$, a value very close to our choice of $0.35~M_{\odot}$, considering the measurement uncertainties for the stellar masses in our samples.

For low-mass M dwarfs ($\rm <0.34~M_{\odot}$ in S21, $\rm <0.35~M_{\odot}$ in our study), our occurrence rates are consistent within 1 $\rm \sigma$, except for the rate obtained for long periods (100 < P < 1000 days) with high-mass planets (100 $<$ \textit{m} sin \textit{i} $<$ 1000 $\rm M_{\oplus}$).
In this range, S21 reports a rate of $16^{+15}_{-9}\%$, with one detection among the 23 stars in their subsample, while we provide an upper limit of 2\% since no planets were detected in this range among the 56 stars in our sample.
This discrepancy stems from the fact that the method used in S21 to compute the occurrence differs from the method we propose.
The computation of the occurrences depends on the detectability rates in the domain probed.
However, such long periods are poorly constrained due to CARMENES’ temporal coverage.
The local detectability rate of their detection in this range (around 60\%) does not correspond to the effective number of stars (Neff) used to compute the rate (around 6.25\%).
In their computation, only the number of detection in the domain is used to validate the frequencies, and not their location and local rate in the domain.
Consequently, the occurrence rate reported by S21 can be overestimated, dominated by the very low detectability rates for long periods.
By applying the local detectability rate to their detection, we obtain a rate consistent with our upper limit.

For more massive M dwarfs, our occurrence frequencies are consistent within 1 $\rm \sigma$ for both short periods (1 < P < 10 days) and long periods (100 < P < 1000 days) across all three planetary mass ranges.
However, for intermediate periods (10 < P < 100 days), our rates are inconsistent for various reasons.
In the low-mass regime (1 $<$ \textit{m} sin \textit{i} $<$ 10 $\rm M_{\oplus}$), S21 reports $210^{+113}_{-81}\%$ with 4 detections, while we report $25^{+12}_{-7}\%$ with 6 detections.
Once again, the 4 detections from S21 are located in the upper left corner of the probed domain. 
The discrepancy between our rates is therefore due to the computation method once again.
The same applies to the intermediate mass range (10 $<$ \textit{m} sin \textit{i} $<$ 100 $\rm M_{\oplus}$), where the detections are again not uniformly distributed.
Finally, in the high-mass regime (100 $<$ \textit{m} sin \textit{i} $<$ 1000 $\rm M_{\oplus}$), S21 reports 2 detections, yielding an occurrence rate of $6^{+4}_{-1}\%$, while we provide an upper limit of 1\% due to the absence of detection in this range.
The 2 detections from S21 correspond to GJ~876 b and c (Karmn J22532–142), for witch the stellar mass reported in Table A1 of S21 is 0.327 $\rm M_{\odot}$.
This mass is below 0.34 $\rm M_{\odot}$, meaning the star was placed in the wrong subsample in S21.
Without this error, our rates are fully consistent. GJ~876 is also part of our sample, and both planets are indeed accounted for in our low-mass star subsample.
In conclusion, aside from this error, the discrepancies between the occurrence rates in our two studies are due to differences in the calculation methods. To avoid raising concerns about errors or going into excessive detail, we have chosen not to delve too deeply into the comparison in the current manuscript.

\subsubsection{HARPS-N occurrence rates obtained in RV}

\cite{pinamonti2022} (hereafter P22) established occurrence frequencies from a sample of 56 massive M dwarfs (M0 - M3) observed with HARPS-N.
In the lowest mass-domain (1 $<$ \textit{m} sin \textit{i} $<$ 10 $\rm M_{\oplus}$), for the shortest periods (1 < P < 10d) they obtain $10.3^{+8.4}_{-3.3}\%$. This is in agreement with the frequency of $11.8^{+7.3}_{-3.3}\%$ we obtained with the sample of massive M dwarfs for the same mass-period ranges.
Concerning the longest periods (10 < P < 100d) they obtain $85^{+5}_{-19}\%$ and we obtain $25^{+12}_{-7}\%$ in this study.
Once again, our frequencies are not in agreement because the frequencies computed by P22 depends on the average value of the detectability in the domain.
Their detections are spread inhomogeneously in the domain, (all in the upper left corner), where the local rate varies between 0.3 to 0.75, far from the average value taken to compute the frequency in that domain. Consequently, with this method, and as explained previously, the frequencies obtained can be overestimated.
In conclusion, the incompatibility between our results in that domain of periods is not due to the sample of targets, but to the method of computation.

\subsection{Dependence on the number of measurements}\label{nmes-dependance}

\begin{figure}
   \centering         
    \includegraphics[width=0.5\textwidth]{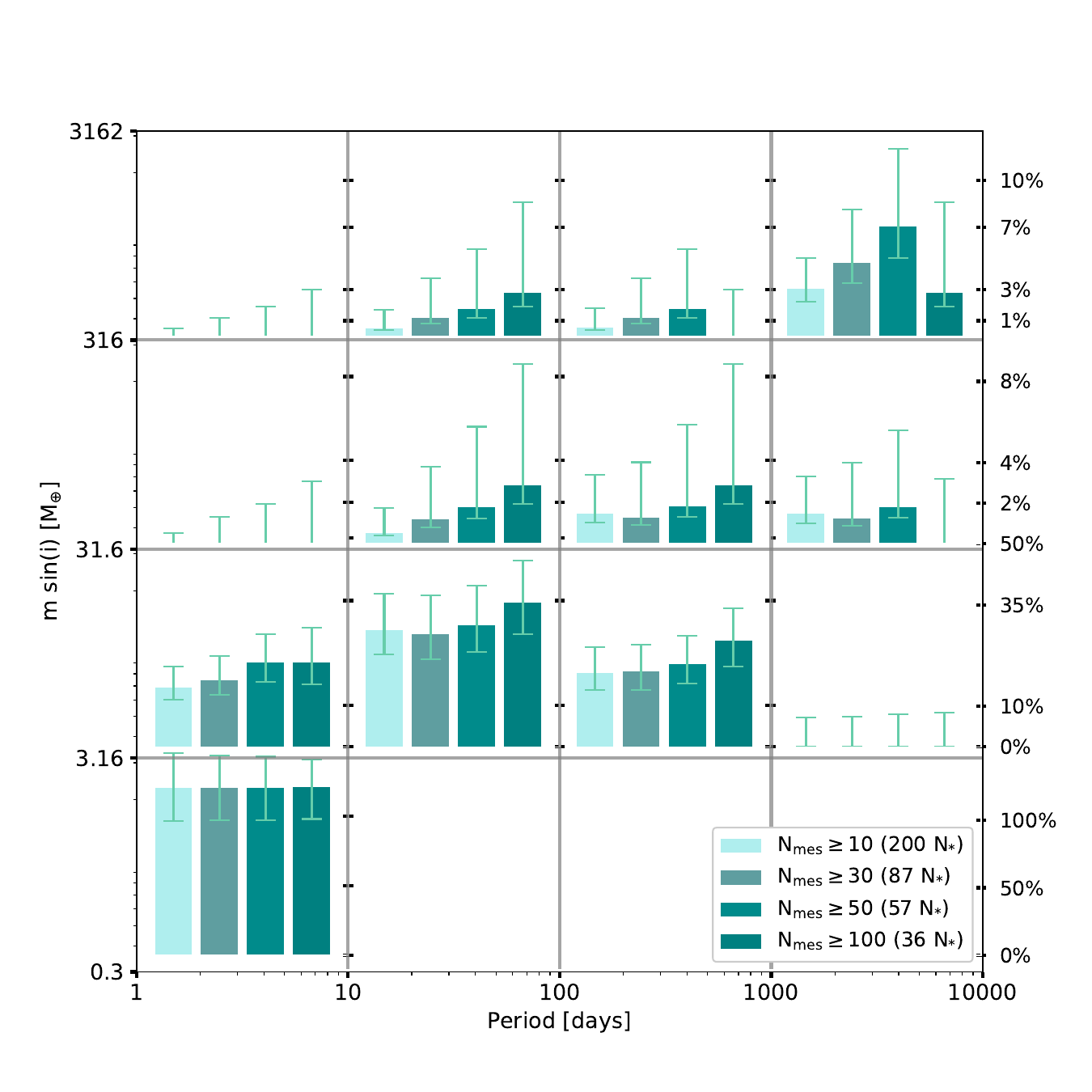}
    \caption{Comparison of occurrence rates obtained as a function of the minimal number of measurements to consider in the mass-period domains probed.}
    \label{statocc_evo}
\end{figure}

This study relies on the sample established in paper I, wherein we decided to consider only M dwarfs observed for at least 10 nights with HARPS.
Comparing the rates obtained by \cite{sabotta2021} on the sample of M dwarfs observed for more than 50 nights with CARMENES raises the question of the impact of the initial sample choice.
We present here the occurence rates obtained on different subsamples of our initial sample by selecting targets based on the number of observations.
Figure~\ref{statocc_evo} displays the rates obtained in the previously probed mass and period domains for four samples (from left to right): the 197 M dwarfs observed for at least 10 nights, the 87 observed for at least 30 nights, the 57 observed for at least 50 nights, and finally the 36 observed for at least 100 nights.

This selection reduces the total number of targets in the probed sample, thereby it alters the weighting of each detection.
Additionally, since the highest detection limits often come from the less extensive RV series, increasing the minimum number of nights in the series purges the detectability map from the upper end.
Specifically, the detection limit at 90\% for the sample of extensively observed M dwarfs (at least 100 nights) corresponds to a much lower mass than that of the M dwarfs in the original sample (observed for at least 10 nights).
These two effects more or less compensate each other, allowing us to distinguish three cases in Fig~\ref{statocc_evo}.
When the total number of stars dominates the ratio, the occurrence rates increase with the minimum number of nights (most of the probed domains).
When the increase in weighting compensates for the decrease in the total number, the occurrence rates plateau, for example in the domain of very low masses at short periods (bottom left corner).
And finally, when the number of detections decreases, the rates fall sharply, as in the domains of long-period Jovian-mass planets (top right corner).
In these domains, the number of detections decreases when considering only the most intensively observed M dwarfs, indicating that systems with such massive cold planets have not been intensively monitored but rather excluded from the HARPS programs.
This is a major bias, as it prevents us from concluding whether there is a correlation, anti-correlation or non-correlation between the internal and external planetary populations, even though this question is now extremely crucial and has not yet been settled \citep[i.e.][]{bonomo2023,Bryan2024}.


\section{Occurrence statistics in Temperate Zone}\label{TZ}


The range of orbital distances, where the conditions permit the presence of liquid water on a planetary surface, is commonly referred to as the Habitable Zone (hereafter HZ, \citealt{huang1959, hart1978}).
Indeed, this study does not address habitability conditions, which are acknowledged to be significantly more complex than the concept discussed herein.
To ensure a clear and unbiased interpretation of this concept, we use a more neutral term such as the Temperate Zone (hereafter TZ).
As the TZ mostly depends on effective temperature of each star, it does not correspond to a fixed domain in period.
To obtain an occurrence rate in TZ of 197 targets, we first convert the detection limits from period dependent to insolation dependent, as in \cite{pinamonti2022}. We then normalise all the detection limits on the inner and outer limits of the TZ, and finally we compute the new occurrence rates in the domain of validity of the model.

\subsection{From periods to insolation}

\begin{figure}
   \centering         
    \includegraphics[width=0.49\textwidth]{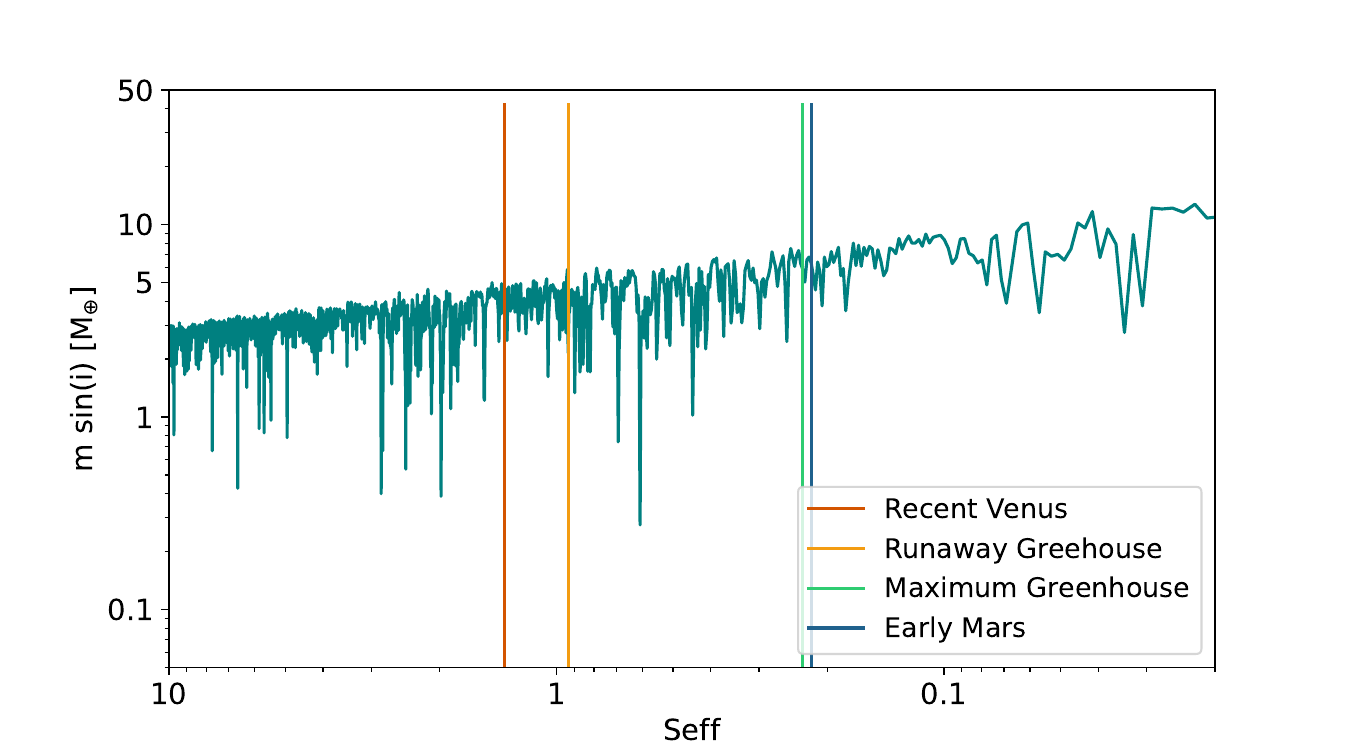}
    \caption{GJ~163 statistical mass limit as a function of insolation. The vertical bars represent the inner (\textit{orange:} Recent Venus, \textit{yellow:} Runaway Greenhouse) and outer (\textit{green:} Maximum Greenhouse, \textit{blue:} Early Mars) boundaries of the HZ defined in \cite{kopparapu2013}.}
    \label{limdet-gj163-insol}
\end{figure}

First, we compute the detection limits from period to distance using the Kepler's third law and then we convert it into insolation:

\begin{equation}
   \rm  Seff = \biggl( \frac{L}{d^2} \biggr)
\label{Seff}
\end{equation}

where the distance d is converted in Effective insolation $\rm S_{eff}$ with the luminosity of the star, L (in L$_{\odot}$).
The luminosity is primarily derived from TESS catalogue\footnote{ExoFOP \url{https://exofop.ipac.caltech.edu/tess/}}.
For the few missing cases, luminosity is computed using the Boltzmann law, with a radius linearly derived from the mass estimated in paper I.
Subsequently, we can transform the detection limits into insolation, as illustred in Fig.~\ref{limdet-gj163-insol}, wich represents the statistical limit of the projected planetary mass as a function of received insolation of GJ~163.

Subsequently, we calculate the values of the inner and outer boundaries ($\rm Seff_i$) of the TZ as defined by \cite{kopparapu2014} for $\rm 1~M_{\oplus}$ (represented by the vertical bars in Fig.~\ref{limdet-gj163-insol}).
To achieve this, we use the i-th set of parameters corresponding to the i-th boundary in the following relation:

\begin{equation}
	\rm Seff_i =  Seff_{i\:\odot} + a_i T + b_i T^2 + c_i T^3 + d_i T^4
\label{Seffbis}
\end{equation}

where T represents the difference between the effective temperature of the star and that of the Sun.
The four sets of parameters used are compiled in Table~\ref{tableparametre}.
Following \cite{kopparapu2014}, we opt for the most conservative boundaries of TZ: Runaway Greenhouse (inner) and Maximum Greenhouse (outer).
After performing this calculation for the 197 targets in our sample, we normalize them by aligning the inner and outer boundaries. 
Consequently, it becomes possible to overlay the 197 detection limits and construct a detectability map similar to the one calculated previously, but this time based on insolation rather than period.
However, it is essential to note that this model is valid only for terrestrial planets, and thus, the planetary mass to consider falls between 0.1 and 10~$\rm M_{\oplus}$.
Therefore, in the following, we calculate the occurrence rate by counting and weighting the detected planets only in the probed TZ domain.
Finally, we obtain the statistical bounds using the same computation as for the planetary occurrence.

\subsection{Results}\label{sect-result-ZT}

\begin{figure}
   \centering         
    \includegraphics[width=0.49\textwidth]{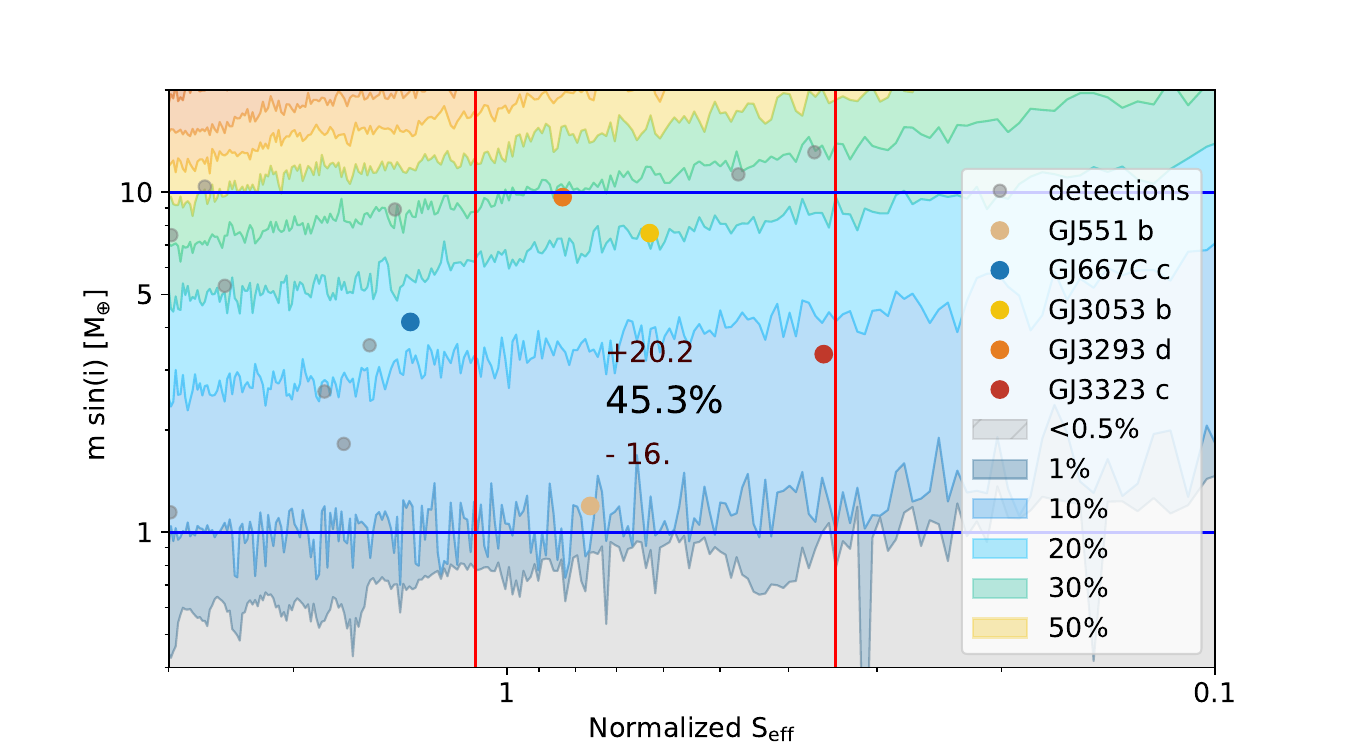}
    \caption{Conservative occurrence rate and planets in the TZ.}
    \label{stat-insol}
\end{figure}

Taking the most conservative boundaries of the TZ, we identify four planets within this zone: GJ~551~b (known as Proxima~b), GJ~3053~b, GJ~3293~d, and GJ~3323~c resulting in an occurrence rate of 45.3\%$\rm ^{+20}_{-16}$, as illustrated in Fig.~\ref{stat-insol}.
This rate is in agreement with other estimation of the planetary occurrence in HZ obtained by RV method (e.g. 41\% \citealp[]{bonfils2013a}).
Our occurrence rate is also $\rm 1\sigma$ consistent with those obtained around M dwarfs using the transit method, with a rate of 15.82\%$\rm ^{+16.60}_{-6.54}$ obtained by \cite{dressing2015} but inconsistent with the 8.58\%$\rm ^{+17.94}_{-8.22}$ obtained more recently by \cite{bergsten2023} on Kepler surveys using models also based on insolation.

It is crucial to emphasize that our rate is derived from a homogeneous analysis of the 197 RV time series, which did not recover 100\% of the planets detected and published to date, and therefore some planets might be missing from our reference sample of detections.
However, it should be noted that if a planet has not been detected, then its signal has not been subtracted from the RV time series, and consequently, this absence is considered in the calculation of the detection limits of this study, thereby affecting the computation of the rate and statistical bounds.
In conclusion, the discrepancy between our rate obtained by RV and those previously obtained by transit is mostly due to the non determination of the radius.
Indeed, working on minimum projected mass, the most massive planets determined in the TZ could be sub-Neptunes rather than tellurics, rendering the applied model invalid, and our rate over-estimated.

\subsection{Planets in TZ}\label{sect-planet-ZT}

\subsubsection{GJ~551~b: Proxima~b}
Although Proxima b is currently considered to be within the Habitable Zone \citep[e.g.][]{faria2022}, it is crucial to note that this determination heavily relies on the effective temperature value.
Without a measurement from TESS, we selected an effective temperature from the five available values on the CDS.
By adopting the most recent value of 3306 K \citep{maldonado2020}, Proxima b falls within the inner zone, receiving excessive insolation.
The other four references suggest values between 2800 and 3000 K, which are more consistent with the star's spectral type (M5.5V). 
Consequently, we selected the second most recent value of 2810 K \citep{kuznetsov2019} for this analysis. 
This case clearly demonstrates the high sensitivity of our study to the effective temperature.

\subsubsection{GJ~3053~b and GJ~3293~d}
These two planets, despite being within the TZ, possess projected masses near the limit of the model's applicability \citep{kopparapu2014}.
In this mass range, the determination of the planet's radius is crucial to distinguish whether they are super-Earths or sub-Neptunes.

\subsubsection{GJ~3323~c}
This planet is on the conservative outer boundary of the TZ (maximum greenhouse).
Consequently, its position is highly sensitive to the effective temperature value adopted from the literature (18 references ranging from 3000 to 3300 K), potentially placing it outside the TZ.

\subsubsection{GJ~667C~c}
Although this star is considered out of the TZ in this study, it is located very close to the conservative inner limit (runaway greenhouse).
The selection of the effective temperature is again crucial: based on the TESS catalog, we have adopted a value of 3650 K. 
However, it is noteworthy that the planet falls on the inner boundary if using the temperature from \citealp[3472 K]{gaidos2014b} and within the TZ when using \citealp[3351 K]{santos2013}.
In addition, recent studies using planetary climate models confirm the presence of GJ~667C~c in the HZ or extremely close to the internal limit (see for example Fig 8 of \cite{turbet2023}).
In the latter case, the occurrence rate of planets in the TZ is 49.6\%$\rm ^{+13}_{-19}$.


\section{Discussion}\label{discu}

\subsection{Intrinsic limitation of the study framework}
We discuss in this section the impact of assumptions chosen to compute this new occurrence rates.

\subsubsection{Limits of correction of activity}

As a first limititation of this study, it is important to point out that 23 series have been corrected for sinusoidal signals attributed to the impact of stellar activity, especially stellar rotation in \cite{mignon23b}.
Subtracting a Keplerian signal does not clearly remove the contribution of stellar activity, especially in the case of long time series, where a change of phase in this signal over time, as well as a variation in its amplitude, have not been taken into consideration.
Given that we considered all series composed at least of 10 nights of observation, it was not feasible to implement a more complex method, such as Gaussian processes, for example.
However, in our sample, 19 stars have more than 150 measurements, and among them, only three (GJ~229, GJ~667C, and GJ~393) exhibited RV signals at varying periods across different seasons.
For these 3 stars, we corrected for two sinusoidal signals with close periods and we may have introduced additional signal into the RV series.
In addition, among the 19 stars, some detection limit masses might also be overestimated because RV time series have not been completely cleaned from stellar signals.
However, since only 3 out of 197 stars are affected by this inappropriate correction, we believe that the results concerning the detection limits of the 197 stars are not significantly compromised.

\subsubsection{Limits of planetary detection method}

Out of the 31 planetary systems detected, 8 are incomplete when compared to the NASA Exoplanet Archive. As an example, GJ~180~c and d, and GJ433 d were detected with complementary data (\cite{tuomi2014} using UVES data, and \cite{feng2020} using VLT2 data). We have verified that all these planets lie below our detection limits and listed them in the Tab.~\ref{tablenodetection} the reference paper, the method and the data of the discovery.

Regarding the impact of these missing planets on the occurrence rates, we cannot include them in our detectability map, as their signals were not removed from the RV series used to determine detection limits. These planets could potentially be detected with more RV data or by using more advanced methods.
With well-sampled RV series, we could compute much lower detection limits, which would likely increase the number of detected planets. However, in the context of our study, this would mean focusing only on the most observed targets. As shown in Fig.~\ref{statocc_evo}, they are already those for which we have detected planets. As a result, this would create a highly biassed sample in the current state of the data.

In conclusion, while the method we have chosen does not recover all previously published planets, it allows us to analyse and process a large number of RV series. We consider this to be the least biassed method possible given the current dataset. The missed planets are all below our detection limits and are therefore accounted for in our statistical boundaries.

\subsubsection{Limits of the binomial model}

In this study, we compute statistical boundaries using a binomial distribution constructed within the surveyed domain, where each draw results in either detection or non-detection.
As explained in Sect.~\ref{secStat}, the distribution reaches its limit in domains with abundant multiple systems. 
Computing an average multiplicity in the proposed domain allows us to artificially increase the number of draws to account for this multiplicity.
Specifically, the conducted experiment involves drawing each star $mult$ times in a draw with replacement, based on the assumption of independent draws.
However, this assumption can be questioned on two distinct aspects. 
Firstly, at the level of multiple system architecture, the presence of a planet with specific mass and period does not necessarily allow the presence of another planet with the same parameter set. 
During the random draw, it may be preferable to narrow down the surveyed zone of the domain by excluding parameter sets prohibited by the first planet, for example.

Secondly, the discovery of one planet has an impact on the discovery of a second. 
Indeed, observations are not conducted blindly, and the RV time series that are better sampled are those selected and strengthened after an initial discovery. 
Our set of RV series is therefore biased by an iterative process of searching for planets in the same system.
While the assumption of independent draws is probably not the optimal approach, manipulating the size of the surveyed domains can render these effects more or less marginal.

\subsubsection{Limits of the new computation of frequencies}

By choosing to weight each detection by its local detectability rate, rather than the average rate over the entire probed domain, we assign significant importance to the planetary parameters (the best fits presented in Tab.~\ref{tabledetection}.

Regarding period uncertainties, the smoothing we applied ensures a uniform rate, as the precision of the map is on the same order of magnitude as the typical error (which is very small in RV measurements).

For uncertainties of the fitted masses, we have performed the following calculation to quantify their impact. We calculated occurrence rates based on 1000 different detection sets, with each set being a random realisation of masses assuming Gaussian errors. We then computed the average value and standard deviation of the resulting occurrence rate distribution. In the vast majority of the probed domains, the standard deviations are much smaller than the statistical bounds we calculated (from 0.01 to few percent), demonstrating that the uncertainties in the masses are not the dominant factor in the occurrence rate calculations; rather, the statistical bounds are. However, it should be noted that for lower mass planets, which are closer to the detection limits, the impact of the uncertainty becomes more significant (22\%).

Finally, it is important to recall that the previous calculation, based on the average detectability rate over the entire probed domain, would imply error bars on the masses of the order of the domain size, which is not realistic for such large parameter spaces.
In conclusion, by using the local detectability rate for each detection, we more accurately capture the distribution of detections within the domain.

\subsubsection{Limits on the occurrence rates at long-term periods}

This study is based on the sample of RV series compiled in paper I where long-term linear and quadratic trends were identified and subtracted from the series.
Some series were corrected for potential signatures of massive companions whose orbits are still too poorly constrained to be characterized, leading to the determination of a minimum period and a minimum mass.
However, the detection limits were calculated based on the residuals of these series, resulting in lowered limits for very long periods (beyond 1000 days).
Consequently, the number of detection at these very long period and the obtained occurrence rate are possibly under-estimated.
It is noteworthy that the vast majority of the massive candidates resulting from the previous study orbit the most massive M dwarfs in our sample.
In conclusion, potentially underestimating the occurrence rate of massive planets at large separations does not fundamentally challenge the observed dichotomy between the least massive and most massive stars concerning this planetary population but may rather tend to reinforce it.

\subsubsection{Limits of the TZ domain model used}

The definition of the TZ we use has been computed for small terrestrial planets orbiting various stellar types, and considering three different planetary masses: 0.1\,M$_\oplus$, 1\,M$_\oplus$ and 10\,M$_\oplus$ \citep{kopparapu2014}.
The stellar type is defined through the effective temperature of the star, from 2600\,K to 7200\,K.
Outside this range the inner edge of the TZ is no longer valid. 
As we do not have access to the real masses of the planets nor the composition of their atmosphere in our sample, we use the TZ estimated for one Earth mass assuming terrestrial composition.
This assumption is commonly use in the literature to fill the lack of data.
\cite{kopparapu2014} propose different inner limits: the "recent Venus", the "runaway greenhouse" and the "moist greenhouse".
In this work, we use the "runaway greenhouse" which is the most conservative definition. 
Moreover, it is interesting to notice that an Earth-like planet (in term of mass and atmospheric composition) is supposed to be tidally locked by assuming a 4.5\,Gyr tidal locking timescale for stars cooler than 4400\,K \citep{kopparapu2014}. In such a scenario, the global circulation in the atmosphere will be drastically different and the inner limit of the TZ will be closer to the star \citep[e.g.][]{yang_stabilizing_2013, kopparapu_habitable_2017}.
This configuration requires 3-D modeling to accurately estimate the inner edge of the TZ but no generic equation as been proposed so far to cover this regime. 

\subsection{Interpretation of the planetary occurrence around M dwarfs}

\subsubsection{The lack of Hot Jupiters}

With no detection of hot Jupiters within the 197 targets, we establish an upper limit of 0.5\% on their occurrence rate around M dwarfs.
This updated upper limit represents the lowest within RV studies around M dwarfs, in agreement with prior studies: < 1.27\% in \cite{endl2006}, < 1\% in \cite{bonfils2013a} and < 3\% in \cite{sabotta2021} (see Fig.~\ref{comp-bonfils}).
This new upper limit accentuates the apparent lack of hot Jupiters around M dwarfs compared to solar-type stars predicted in \cite{endl2006}.
Indeed, occurrence rates of hot Jupiters detected via RV around FGK-type stars obtained so far are slightly higher: e.g. 1.5\% in \cite{cumming2008}, $1.2 \pm 0.4$ in \cite{wright2012}, and more recently 0.84\% in \cite{wittenmyer2020}.
Without any detection of a hot Jupiter around an M dwarf in our probed sample of 197 M dwarfs, this upper limit only depends on the total volume of the sample.
Thanks to the extensive volumes surveyed and the high sensitivity to hot Jupiters, the transit technique emerges as the most adept in refining the population contrast between M dwarfs and FGK stars.
Notably, a recent study by \cite{gan2023}, based on the TESS survey of M dwarfs, reported a rate of $ \rm 0.27 \pm 0.09 \%$, slightly lower than the rates of $0.43^{+0.07}_{-0.06}\%$ by \cite{masuda2017} and $0.57^{+0.14}_{-0.12}\%$ by \cite{petigura2018}, both based on FGK-type samples using the transit method.
Although still consistent within 2 sigma, these frequencies obtained on transit surveys will need to be refined to account for the observed lack of hot Jupiters around M dwarfs as the RV method alone will not provide a large enough sample of M dwarfs to establish statistically significant conclusions on this matter.

\subsubsection{Dependence of planetary population on the central mass}

\begin{figure}
   \centering         
    \includegraphics[width=0.95\linewidth]{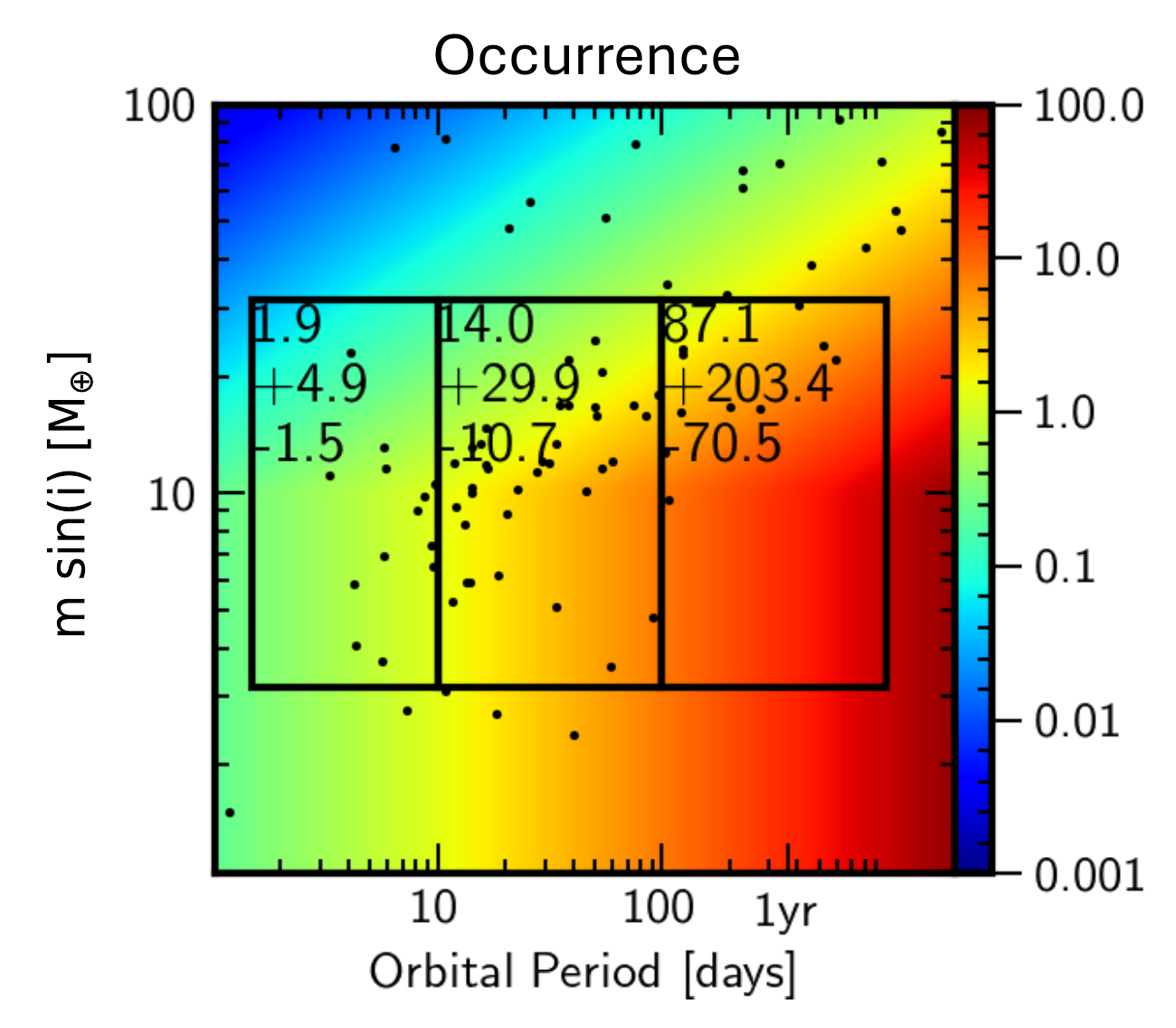}
    \caption{Occurrence of low-mass planets around FGK stars based on \citet{mayor2011} obtained using EPOS \citep{mulders2018}. The marked bins are the ones also used in Figs. \ref{statocc}, \ref{statocc_sup35}, and \ref{statocc_inf35} and occurrence is indicated in percent and color-coded in logarithmic mass against period space.}
    \label{mayor_epos}
\end{figure}

It is of particular interest to probe the dependency of planetary occurrence with stellar mass as this dimension can reveal new constraints on planet formation. The most accessible, first-order analysis involves the occurrence of planets as a function of stellar mass. From transit missions, a decreasing occurrence with increasing stellar mass was found \citep{dressing2013,mulders2015,dressing2015,hsu2020}. Recently, based on updated stellar parameters, the conclusion was questioned again \citep{bergsten2023}.
Although based on a smaller sample of stars, our results based on the radial velocity method can help to complement these findings. 

There is not significant difference in planetary occurrence for the planets with \textit{m} sin \textit{i} from 3.16 to 31.6~M$_\oplus$ at each respective period range. The lowest mass bin (Figs. \ref{statocc}, \ref{statocc_sup35}, and \ref{statocc_inf35}) with planets below 3.16~M$_\oplus$ is inaccessible for the more massive stars. At larger planetary masses (<100~M$_\oplus$), there is evidence that more planets exist around low-mass stars but the number of planets is still limited.

The occurrence for FGK stars based on the RV method can be estimated from the \citet{mayor2011} sample using EPOS \citep{mulders2018} and a broken power-law in orbital period and mass space. To not be influenced by a potentially different giant planet population, we constrained the fit to the planets with \textit{m} sin \textit{i} $<$ 100 $\rm M_{\oplus}$ and investigate the super-Earth to sub-Neptune mass bin ((3.2 $<$ \textit{m} sin \textit{i} $<$ 31.6 $\rm M_{\oplus}$)) shown in Fig. \ref{mayor_epos}. Since this is a regime which is challenging to assess around Solar-mass stars, useful constraints can only be obtained for planets with $P<100$\,days. In the innermost period bin ($1.5<P<10$\,days), we find an occurrence of $1.9^{+4.9}_{-1.5}\%$ which is significantly smaller than around M stars as reported here. This observation was also made by \citet{sabotta2021} and was discussed in \citet{schlecker2022}. They hypothesized that a migration trap which depends on the disk temperature (and therefore also on stellar properties) could produce this outcome.

For the bin extending from $10<P<100$\,days, the EPOS analysis results in an occurrence of $14^{+29.9}_{-10.7}$ which is, given the large uncertainty, in agreement with the same mass-period region around M stars. Given that transit observations are mainly probing the innermost orbital periods, the decrease in planet occurrence with increasing stellar mass might be mainly a close-orbit effect and we can not provide evidence with radial velocity surveys that it also holds at larger orbital periods.

\subsubsection{Comparison with planetary population synthesis}
Theoretical models so-far fail to comprehensively explain the correct stellar-mass dependent planet distribution. For the aforementioned overall occurrence trend with stellar mass, \citet{burn2021} predict a slight increase of occurrence from FGK to early M before it decreases again due to solid-mass limitation. The reason for the lower occurrence around FGK stars is the more common presence of massive giant planets and their gravitational perturbations of lower-mass planets. Further effects are growth to larger mass bins as well as differences in minimum spacing between two planets if they become more massive due to their more extended Hill sphere.

Another effect could be the halting of the pebble flux to the inner system which is not included in that work \citep{lambrechts2019,mulders2021}. However, if the decrease in planetary occurrence with stellar mass in planetary occurrence is as large as reported by transit observations, the observed occurrence of giants around every fourth FGK star is insufficient to explain the full effect. Furthermore, if the effect acts, it should also produce a strong super-Earth to cold-Jupiter anti-correlation which is not convincingly observed yet in studies \citep{barbato2018,bryan2019} and unprovable in this study due to the observational bias highlighted in Sec.~\ref{nmes-dependance}.

\begin{figure}
   \centering         
    \includegraphics[width=0.49\textwidth]{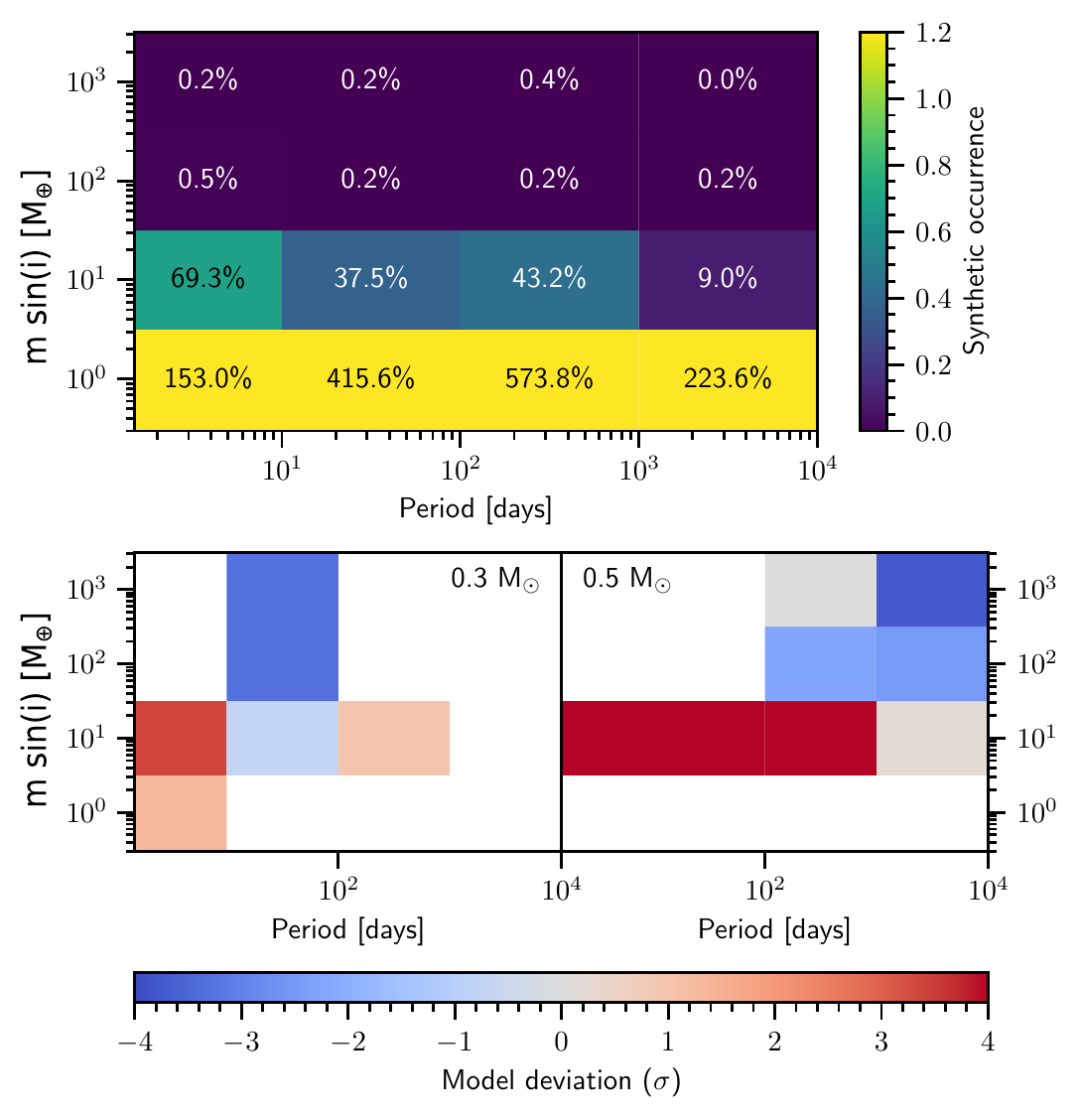}
    \caption{Occurrence rates of theoretically calculated planets by \citet{burn2021}. The same bins as in Fig.~\ref{statocc}, \ref{statocc_sup35}, and \ref{statocc_inf35} are used. Values of $\sin i$ are randomly assigned to the synthetic systems following the expected distribution for randomly oriented systems \citep{schlecker2022}. The top panel shows the synthetic occurrence for the combined discrete stellar masses of 0.3\,M$_{\odot}$ and 0.5\,M$_{\odot}$. The bottom panels show the deviation from observations measured in standard deviations $(f_{\rm p, synthetic} - f_{\rm p})/s_{f_{\rm p}}$ for the indicated stellar masses compared to the observational data from Figs.~\ref{statocc_inf35} (left) and \ref{statocc_sup35} (right). Where observational upper limits are in agreement with theory, the box is left white.}
    \label{synthetic}
\end{figure}

Despite the theoretical uncertainties, to test planet formation models, it is interesting to compare synthetic planetary populations against the here derived occurrence rates. Fig. \ref{synthetic} shows the occurrence of synthetic planets based on the modeling by \citet{burn2021}. Their model includes comprehensive disk evolution, growth of planets by planetesimal accretion, N-body interactions, orbital migration, and an evolving planetary envelope and interior resolved in one dimension. 

The general conclusion from the comparison is that there are too few giant planets forming in the synthetic models around M dwarfs while small planets are over-produced. The over-production is more significant for the simulation around 0.5\,M$_\odot$ stars compared to the late M dwarfs. However, we note that a similar over-production of model planets is present for Solar mass stars.

Potential model components which should be adjusted are to include pebble accretion which might promote more growth to giant planets while also only including initial seeds at locations and masses which are realistic \citep{voelkel2021}. Furthermore, the radial planetesimal distribution could be less steep or show different features compared to their assumed steep profile (exponent of -1.5). As discussed in detail in \citep{schlecker2022}, the innermost population of planets should be revised around more massive Stars.

\subsubsection{1D vs 3D atmospheric models}

The definition of the TZ given by \cite{kopparapu2014} is largely used thanks to the useful and generic equation they provide. However, they used a 1-D climate model \citep{kopparapu2013} for their calculation. Some studies \citep[e.g.][]{yang_stabilizing_2013, kopparapu_inner_2016,kopparapu_habitable_2017, turbet_water_2023} showed that intrinsically 3-D processes such as the dynamics and the clouds strongly affect the position of the inner edge of the TZ. 

It could be interesting to use a more accurate definition of the TZ, based on 3-D climate modeling, to revisit the occurrences proposed in this work. In the same way, considering atmospheric compositions consistent with the mass of the planet, for example (e.g., Earth-like for masses around one earth mass, and H$_2$ dominated for masses higher than 5 Earth masses), could greatly improve our estimations. This is the aim of a future work.


\section{Conclusion}\label{conclu}

Built upon a prior homogeneous analysis of 197 M dwarfs, our study enables us to present the most precise occurrence rates in RV studies to date. 
We confirm the prevalence of low-mass, short-period planets, with a rate of 120\%$\rm ^{+25}_{-24}$ within a 10-day orbital period for planets under $\rm 3~M_{\oplus}$, and we begin to explore planets with projected masses below $\rm 1~M_{\oplus}$.
The obtained rates are consistent with those from other RV studies, and we propose a new rate for planets at very large separations facilitated by the unique longevity of HARPS.
By dividing our sample of M dwarfs by stellar mass, we reveal the expected dichotomy between populations of massive planets at large separations, none of them being detected around the less massive M dwarfs.
In addition, extending this study to the insolation calculation yields an occurrence rate of 45.3\%$\rm ^{+20}_{-16}$ in the TZ of M dwarfs, suggesting a large number of potential targets for this domain of research, given the number of M dwarfs in the solar neighbourhood.
Beyond these main results, this study highlight the limitations faced when computing occurence statistics, even when considering a sample that is analysed in an homogeneous way.
Finally, this study also shows the impact of sampling and the choices made over the last 20 years in M dwarf monitoring programmes.


\begin{acknowledgements} 
This research has made use of the VizieR catalogue access tool, CDS, Strasbourg, France. The original description of the VizieR service was published in A\&AS 143, 23
This work has been supported by a grant from LabEx OSUG@2020 (Investissements d'avenir - ANR10LABX56) 
This research has made use of the NASA Exoplanet Archive, which is operated by the California Institute of Technology, under contract with the National Aeronautics and Space Administration under the Exoplanet Exploration Program.
This research has made use of the Exoplanet Follow-up Observation Program (ExoFOP; DOI: 10.26134/ExoFOP5) website, which is operated by the California Institute of Technology, under contract with the National Aeronautics and Space Administration under the Exoplanet Exploration Program.
This work has been carried out within the framework of the NCCR PlanetS supported by the Swiss National Science Foundation.
NCS acknowledges funding by the European Union (ERC, FIERCE, 101052347). Views and opinions expressed are however those of the author(s) only and do not necessarily reflect those of the European Union or the European Research Council. Neither the European Union nor the granting authority can be held responsible for them. This work was supported by FCT - Fundação para a Ciência e a Tecnologia through national funds and by FEDER through COMPETE2020 - Programa Operacional Competitividade e Internacionalização by these grants: UIDB/04434/2020; UIDP/04434/2020. N.A-D. acknowledges the support of FONDECYT project 1240916. X.D acknowledge from support by the French National Research Agency in the framework of the Investissements d’Avenir program (ANR-15-IDEX-02), through the funding of the "Origin of Life" project of the Univ. Grenoble-Alpes.
GC acknowledge the financial support of the SNSF (grant number: 200021\_197176 and 200020\_215760).
\end{acknowledgements}

\bibliographystyle{aa}
\bibliography{library}

\begin{appendix}
\onecolumn

\section{Tables} \label{append1}

\begin{longtable}{r|c|r|c|r|c}
\caption{\label{progidtable}Program Id and PIs} \\
PI & Prog. Id &PI & Prog. Id &PI & Prog. Id \\ \hline 
\hline
\endhead

\multirow{6}{*}{Bonfils}&198.C-0873 &\multirow{14}{*}{Lo Curto}&196.C-1006&\multirow{2}{*}{Udry} & 192.C-0852\\
& 191.C-0873 &  & 099.C-0458 & & 183.C-0972 \\ \cline{5-6}
 & 183.C-0437 & & 098.C-0366 & \multirow{5}{*}{Lagrange} & 192.C-0224 \\
 & 180.C-0886 & & 096.C-0460 & &  099.C-0205 \\
  & 082.C-0718 & & 095.C-0551 & & 098.C-0739 \\
  & 1102.C-0339 & &  093.C-0409 & & 089.C-0739 \\ \cline{1-2}
  \multirow{2}{*}{ Anglada-Escude} & 191.C-0505 & & 091.C-0034 & &0104.C-0418 \\ \cline{5-6}
 & 096.C-0082 & & 090.C-0421& Lannier &  097.C-0864 \\ \cline{5-6} \cline{1-2}
 Astudillo Defru & 100.C-0884 & & 089.C-0732 & \multirow{3}{*}{Diaz} & 098.C-0518 \\ \cline{1-2}
 Poretti &185.D-0056 & & 087.C-0831& & 096.C-0499 \\ \cline{1-2}
 Lachaume & 089.C-0006 & & 085.C-0019 & & 0100.C-0487 \\ \cline{1-2} \cline{5-6}
 \multirow{2}{*}{Haswell} & 097.C-0390 & &  0102.C-0558 & \multirow{2}{*}{Ruiz} & 090.C-0395 \\
 &  096.C-0876 & & 0103.C-0432 & & 084.C-0228 \\ \cline{1-2} \cline{5-6}
 Kuerster & 097.C-0090 & &  0101.C-0379 & Santos & 086.C-0284 \\ \cline{1-2} \cline{5-6}
 \multirow{2}{*}{Mayor} & 077.C-0364 & & 0100.C-0097 & Pepe & 089.C-0050 \\ \cline{3-6}
 &  072.C-0488 & Dall & 078.D-0245 & Hebrard & 078.C-0044 \\ \hline
 Albrecht & 095.C-0718  &  Sterzik & 079.C-0463 & Galland & 076.C-0279 \\ \hline
 \multirow{3}{*}{Guenther} & 076.C-0010 & Melo & 076.C-0155 & Jeffers & 0104.C-0863 \\ \cline{3-6}
 & 075.C-0202 & Robichon & 074.C-0364 & Trifonov & 0100.C-0414 \\ \cline{3-6}
 & 074.C-0037 & Berdinas &  0101.D-0494 & Debernardi & 075.D-0614 \\
\end{longtable}

\begin{longtable}{  r || c|  c|  c| c}
	\caption{\label{tabledetection}Published planets used to build statistics occurrence.} \\
 	Name & Period [d] & $\rm m sin(i)$ [$\rm M_J$] &  $\rm m sin(i)$ [M$_{\oplus}$]  & Reference \\
 	\hline
 	\hline
 	\endhead
 	
 	GJ~54.1~c &  3.0599$\pm$0.0002 & 0.0031$\pm$0.0004 & 0.98$\pm$0.14 & \cite{astudillo2017B} \\
	d & 4.6566$\pm$0.0003 & 0.0034$\pm$0.0005 & 1.14$\pm$0.17 & \cite{astudillo2017B} \\
 	b &  2.02101$\pm$0.00005 &  0.0024$\pm$0.0004 & 0.75$\pm$0.13 &  \cite{astudillo2017B} \\
 	\hline
 	GJ~163~b & 8.6300$\pm$0.0003 & 0.033$\pm$0.002 & 10.6$\pm$0.6 & \cite{bonfils2013b} \\
	c &  25.608$\pm$0.006 & 0.021$\pm$0.003 & 6.8$\pm$0.9 & \cite{bonfils2013b} \\
 	d & 642.15$\pm$1.65 & 0.093$\pm$0.009 & 29.4$\pm$2.9 & \cite{bonfils2013b}\\
 	\hline
 	GJ~176~b & 8.7754$\pm$0.0002 & 0.029$\pm$0.004 & 9.06$\pm$1.37 & \cite{rosenthal2021} \\
 	\hline
 	GJ~179~b & 2321$\pm$500 & 0.752$\pm$0.041 & 239$\pm$13 & \cite{rosenthal2021} \\
 	\hline
 	GJ~180~b & 17.124$\pm$0.006 & 0.020$\pm$0.002 & 6.49$\pm$0.68 & \cite{feng2020} \\
 	\hline
 	GJ~221~c & 125.44$\pm$0.09 & 0.17$\pm$0.03 & 53$\pm$8 & \cite{locurto2013} \\
 	b & 3.87$\pm$0.01 & 0.027$\pm$0.004 & 8.5$\pm$1.3 & \cite{locurto2013} \\
 	\hline
 	GJ~229~c & 121.49$\pm$0.07 & 0.023$\pm$0.004 & 7.268$\pm$1.256 & \cite{feng2020} \\
 	b & 528.17$\pm$6.7 & 0.0267$\pm$0.0064 & 8.478$\pm$2.033 & \cite{feng2020} \\
 	\hline
 	GJ~273~b & 18.661$\pm$0.004 & 0.0091+0.0009 & 	2.89$\pm$0.27 & \cite{astudillo2017A} \\
 	c & 4.7230$\pm$0.0002 & 0.0037$\pm$0.0005 & 1.18$\pm$0.16 & \cite{astudillo2017A} \\
 	\hline
 	GJ~317~b & 695.660$\pm$0.355 & 1.753$\pm$0.058 & 557.1$\pm$18.3 & \cite{feng2020} \\
 	c & 3948$\pm$113 & 1.4 & 420 & \cite{mignon23b} \\
 	\hline
  	GJ~393~b & 7.0223$\pm$0.0008 & 0.00538$\pm$0.00076 & 1.71$\pm$0.24 & \cite{amado2021} \\
  	\hline
	GJ~433~b & 7.3721$\pm$0.0006 & 0.0190$\pm$0.0019 & 6.043$\pm$0.597 & \cite{feng2020} \\
	\hline
	GJ~447~b & 9.861$\pm$0.002 & 0.00440$\pm$0.00066 & 1.40$\pm$0.21 & \cite{bonfils2018} \\
	\hline
	GJ~480~b & 9.548$\pm$0.005 & 0.0415$\pm$0.0053 & 13.2$\pm$1.7 & \cite{feng2020} \\
	\hline
	GJ~536~b & 9.7089$\pm$0.0004 & 0.0205$\pm$0.0022 & 6.52$\pm$0.69 & \cite{trifonov2018} \\
	\hline
	GJ~551~b & 11.187$\pm$0.001 & 0.0040$\pm$0.0006 & 1.27$\pm$0.19 & \cite{anglada2016} \\
	\hline
	GJ~581~b & 5.36856$\pm$0.00006 & 0.0511$\pm$0.0019 & 16.2$\pm$0.6 & \cite{rosenthal2021} \\
	c & 12.914$\pm$0.001 & 0.0159$\pm$0.0022 & 5.05$\pm$0.70 & \cite{rosenthal2021} \\
	e & 3.1489$\pm$0.0001 & 0.00521$\pm$0.00076 & 1.657$\pm$0.024 & \cite{trifonov2018} \\
	\hline
	GJ~628~d & 217.914$\pm$0.305 & 0.0242$\pm$0.0035 & 7.70$\pm$1.12 & \cite{astudillo2017A} \\
	c & 17.865$\pm$0.003 & 0.0107$\pm$0.0014 & 3.41$\pm$0.43 & \cite{astudillo2017A} \\
	b & 4.8872$\pm$0.0003 & 0.0060$\pm$0.0008 & 1.91$\pm$0.26 & \cite{astudillo2017A} \\
	\hline
	GJ~667C~b & 7.2005$\pm$0.0001 & 0.018$\pm$0.001 & 5.6$\pm$0.3 & \cite{robertson2014} \\
	d & 90.96$\pm$0.03 & 0.0160$\pm$0.0057 & 4.8$\pm$1.7 & \cite{anglada2013} \\
	c & 28.151$\pm$0.005 & 0.013$\pm$0.002 & 4.1$\pm$0.6 & \cite{robertson2014} \\
	\hline
	GJ~674~b & 4.69507$\pm$0.00002 & 0.035 & 11.09 & \cite{bonfils2007} \\
	\hline
	GJ~676A~b & 1050.3$\pm$1.2 & 4.95$\pm$0.31 & 1570$\pm$100 & \cite{anglada2012} \\
	\hline
	GJ~752A~b & 105.911$\pm$0.109 & 0.03843$\pm$0.00330 & 12.214$\pm$1.050 & \cite{kaminski2018} \\
	\hline
	GJ~832~b & 3684$\pm$97 & 0.68$\pm$0.09 & 216$\pm$29 & \cite{wittenmyer2014} \\
	\hline
	GJ~849~b & 1940$\pm$12 & 0.891$\pm$0.036 & 283$\pm$11 & \cite{rosenthal2021} \\
	\hline
	GJ~876~b & 61.1166$\pm$0.0086 & 2.2756$\pm$0.0045 & 723.2$\pm$1.4 & \cite{rivera2010} \\
	c & 30.2310$\pm$0.0002 & 0.714$\pm$0.004 & 226.98$\pm$1.23 & \cite{rivera2010} \\
	d & 1.938$\pm$0.001 & 0.0254$\pm$0.0010 & 8.06$\pm$0.31 & \cite{trifonov2018} \\
	\hline
	GJ~3053~b &  24.84$\pm$0.03 & 0.0192$\pm$0.0015 & 6.09$\pm$0.48 & \cite{lillo2020} \\
	c & 3.7816$\pm$0.0008 & 0.00538$\pm$0.00057 & 1.71$\pm$0.18 & \cite{lillo2020} \\
	\hline
	GJ~3293~b & 30.611$\pm$0.006 & 0.07406$\pm$0.00277 & 23.54$\pm$0.88 & \cite{astudillo2017A} \\
	c & 123.4$\pm$0.1 & 0.06636$\pm$0.00390 & 21.09$\pm$1.24 & \cite{astudillo2017A} \\
	d & 48.16$\pm$0.03 & 0.0239$\pm$0.0033 & 7.60$\pm$1.05 & \cite{astudillo2017A} \\
	\hline
	GJ~3323~b & 5.3636$\pm$0.0007 & 0.0064$\pm$0.0008 & 2.02$\pm$0.26 & \cite{astudillo2017A} \\
	c & 40.36$\pm$0.07 & 0.00727$\pm$0.00157 & 2.31$\pm$0.50 & \cite{astudillo2017A} \\
	\hline
	GJ~3341~b & 14.208$\pm$0.005 & 0.0208$\pm$0.0003 & 6.6$\pm$0.1 & \cite{astudillo2015harps} \\
	\hline
	GJ~3634~b & 2.6458$\pm$0.0002 & 0.022$\pm$0.003 & 7.0$\pm$0.9 & \cite{bonfils2011} \\
  	\hline
  	GJ~9018~b & 15.79$\pm$0.02 & 0.0409$\pm$0.0129 & 13.0$\pm$6.6 & \cite{tuomi2014} \\

\end{longtable}

\begin{longtable}{  r || c|  c|  c| c | c}
	\caption{\label{tablenodetection}Additional planets published in planetary systems.} \\
 	Name & $Period_p$ & $\rm m sin(i)_p$ &  Detection   & Reference  & Data used \\
         & [d] & [M$_{\oplus}$] & limit [M$_{\oplus}$] & paper & \\
 	\hline
 	\hline
 	\endhead
 	
 	GJ~180~c &  24.3 & 6.4 & 7.2 & \cite{tuomi2014} & HARPS + UVES\\
	d & 106 & 7.6 & 10.5 & \cite{tuomi2014} & HARPS + UVES \\
        \hline
 	GJ~433~c &  5100 &  32 & 56.7 &  \cite{tuomi2014} & HARPS + UVES \\
        d & 36 & 5 & 8.2 & \cite{feng2020} & HARPS + VLT2 \\
 	\hline
   	GJ~676A~c & 13000 & 4288 & - & \cite{sahlmann16} & Astrometry + RV \\
         & & & & \cite{feng22} & \\
 	$\rm d^*$ & 3.6 & 3.8 & 7.3 & \cite{anglada2012} & HARPS \\
        $\rm e^*$ & 35 & 6.7 & 18.4 & \cite{anglada2012} & HARPS \\
 	\hline
   	GJ~849~c & 5990 & 315 & $>10^3$ & \cite{montet14} & RV + direct imaging \\
        & & & &\cite{pinamonti23} & \\
        \hline
   	GJ~876~e & 124 & 14.6 & 18.5 & \cite{rivera2010} & Keck/Hires \\
        & & & & \cite{montet14} & \\
        \hline
        GJ~3053~c & 3.8 & 1.7 & 1.9 & \cite{ment19} & transit + RV \\
        \hline
        GJ~3293~e & 13 & 3 & 4.1 & \cite{astudillo2017B} & HARPS \\
        \hline

\end{longtable}
\tablefoot{Additional planets not detected in the planetary systems used and detailed in Tab.~\ref{tabledetection}. $\rm Period_p$ and $M.sin(i)_p$ are the planetary parameters published in the reference paper, Detection limits are the average value of the detection limit in the domain of period of the period published computed in this study, and the Data used correspond to the method/data used to detect the additionnal planet in the system. $^*$GJ~676~d and e are found using the same HARPS data set as \cite{anglada2012} obtained before 2011, but not with the new full data set.}

\begin{longtable}{   l | c || c | c|  c|  c}
\caption{\label{tableoccu_comp}Comparison of occurrence rates} \\

Planetary Mass range [$\rm M_{\oplus}$] & Ref & 1 - 10 d & 10 - 100 d & 100 - 1000 d & 1000 - 10000 d \\
\hline
\hline
\endhead

\multirow{3}{6em}{100 - 1000} & S21  & < 3 & $\rm 4^{+3}_{-2}$ & $\rm 5^{+4}_{-3}$ & - \\
& M24 & < 0.5 & $\rm 1^{+1}_{-0.3}$ & $\rm 1^{+1.5}_{-0.3}$ & $\rm 3^{+2}_{-1}$ \\
& B13 & < 1 & $\rm 2^{+3}_{-1}$ & < 1 & $\rm 4^{+5}_{-1}$ \\ \hline

\multirow{3}{6em}{10 - 100} & S21 & $\rm 3^{+3}_{-2}$ & $\rm 10^{+5}_{-4}$ & $\rm 14^{+10}_{-7}$ & - \\
& M24 & $\rm 3^{+2}_{-1}$ & $\rm 3^{+3}_{-1}$ & $\rm 13^{+7}_{-3}$ & < 3 \\
& B13 & $\rm 3^{+4}_{-1}$ & < 2 & < 4 & < 12\\ \hline

\multirow{3}{6em}{1 - 10} & S21 & $\rm 59^{+20}_{-17}$ & $\rm 97^{+42}_{-33}$ & - & -  \\
& M24 & $\rm 89^{+18}_{-15}$ & $\rm 90^{+21}_{-18}$ & $\rm 6^{+5}_{-2}$ & - \\
& B13 & $\rm 36^{+25}_{-10}$ & $\rm 52^{+50}_{-16}$ & -  & - \\

\end{longtable}
\tablefoot{Planetary occurrences rates in percent from this work (M24), from \cite{sabotta2021} (S21) and from \cite{bonfils2013a} (B13).}

\begin{longtable}{   l || c | c|  c|  c| c}
\caption{\label{tableparametre}Parameters from \cite{kopparapu2014}} \\
 	Model & Seff & a & b & c  & d \\
 	\hline
 	\hline
 	\endhead
Recent Venus & 1.776 & $2.136\times10^{-4}$ & $2.533\times10^{-8}$ & $-1.332\times10^{-11}$ & $-3.097\times10^{-15}$ \\
Runaway Greenhouse & 1.107 & $1.332\times10^{-4}$ & $1.58\times10^{-8}$ & $-8.308\times10^{-12}$ & $-1.931\times10^{-15}$ \\
Maximum Greenhouse &0.356 & $6.171\times10^{-5}$ & $1.698\times10^{-9}$ & $-3.198\times10^{-12}$ & $-5.575\times10^{-16}$ \\
Early Mars & 0.32 & $5.4513\times10^{-5}$ & $1.526\times10^{-9}$ & $-2.874\times10^{-12}$ & $-5.011\times10^{-16}$ \\
\end{longtable}
\tablefoot{Parameters used for each model of inner and outer limit of the HZ}



\section{Graphes}\label{append2}

\begin{figure}[h!]
   \centering         
    \includegraphics[width=0.89\textwidth]{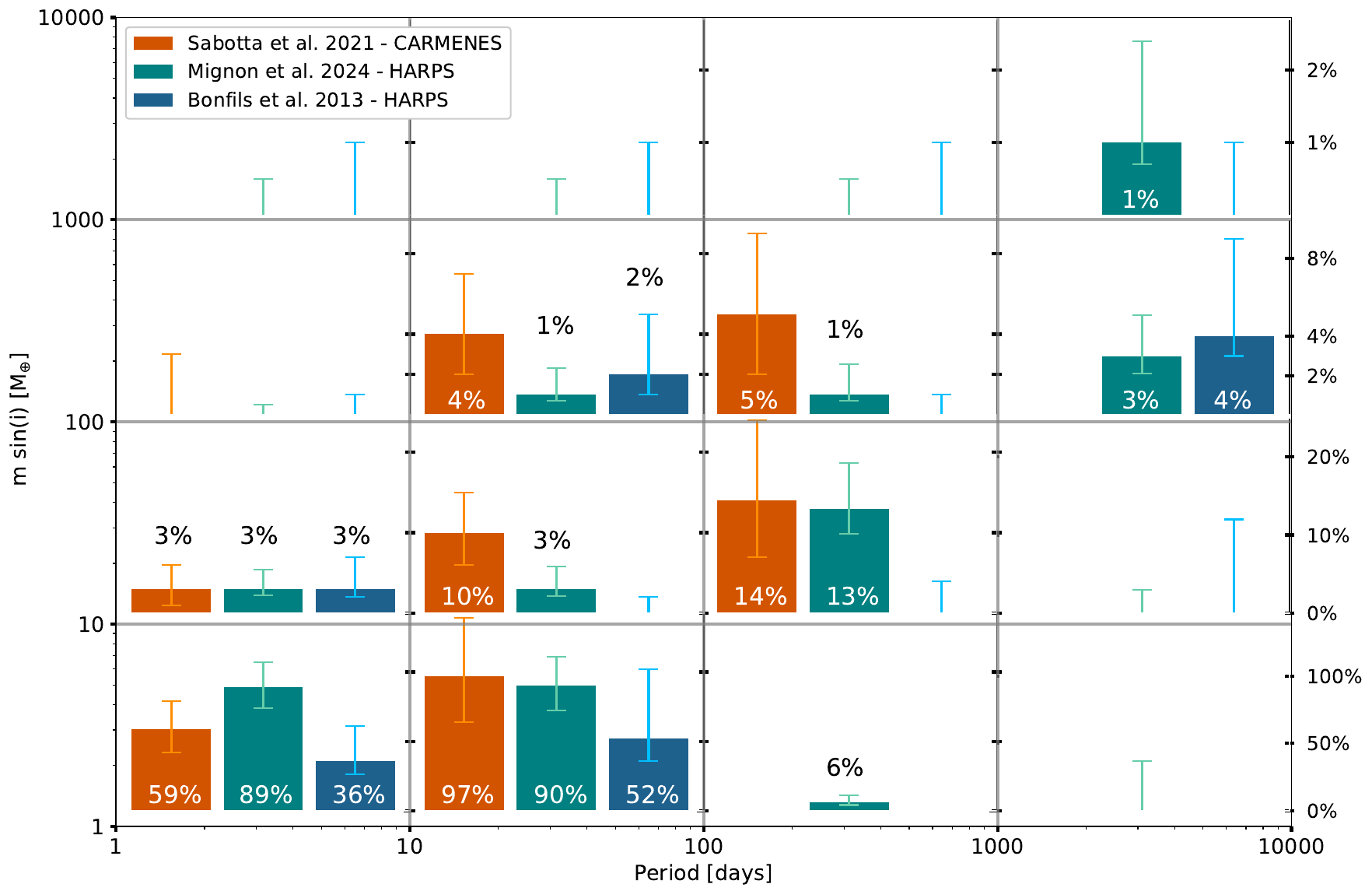}
    \caption{Planetary occurrence rates from \cite{sabotta2021} (orange), those derived from this study (green) and those from \cite{bonfils2013a} (blue), compared for each range of projected planetary mass and period.}
    \label{comp-bonfils}
\end{figure}

\end{appendix}

\end{document}